\documentclass[]{aa}

\usepackage{graphicx}
\usepackage{xcolor}
\usepackage{txfonts}
\usepackage{hyperref}
\usepackage{lscape}
\usepackage{tablefootnote}
\usepackage{placeins}

\begin{document} 

   \title{Nascent chemical complexity in prestellar core IRAS 16293E: complex organics and deuterated methanol }\titlerunning{Nascent chemical complexity in prestellar core IRAS 16293E}

   \author{S. Scibelli\inst{1}
          \and
          M. N. Drozdovskaya\inst{2}
          \and
          P. Caselli\inst{3}
          \and
          J. Ferrer Asensio\inst{4}
          \and
          B. Kulterer\inst{5}
          \and \\
          S. Spezzano\inst{3} 
          \and
          Y. Lin\inst{3}
          \and
          Y. Shirley\inst{6} 
          }

   \institute{National Radio Astronomy Observatory, 520 Edgemont Road, Charlottesville, VA 22903, USA\\
              \email{sscibell@nrao.edu}
              \thanks{Jansky Fellow of the National Radio Astronomy Observatory.}
        \and
            Physikalisch-Meteorologisches Observatorium Davos und Weltstrahlungszentrum (PMOD/WRC), Dorfstrasse 33, CH-7260, Davos Dorf, Switzerland\\
            \email{maria.drozdovskaya.space@gmail.com}
        \and
              Max-Planck-Institut für extraterrestrische Physik, Giessenbachstrasse 1, 85748 \indent Garching, Germany 
        \and 
            RIKEN Cluster for Pioneering Research, Wako-shi, Saitama, 351-0106, Japan
        \and 
            Center for Astrophysics | Harvard \& Smithsonian, 60 Garden St., Cambridge, MA 02138, USA
        \and 
            Steward Observatory,  University of Arizona, 933 North Cherry Avenue,
            Tucson, AZ 85721, USA
                \\
             }
             
\date{Received XX; accepted XX}

  \abstract
  {Prestellar cores represent early sites of low-mass ($M$ $\leq$ few M$_\odot$) star and planet formation and provide insight into initial chemical conditions of complex organic molecules (COMs). Deuterated COMs trace the degree of molecular inheritance and/or reprocessing, as high deuteration in protostellar systems suggests COMs forming during the prestellar stage when deuteration is enhanced.}
  {Within the L1689N molecular cloud, the prestellar core IRAS 16293E sits $90\arcsec$ eastward of the chemically-rich IRAS 16293-2422 A and B protostellar system. A unique view of star formation inside a common natal cloud, IRAS\,16293A, B, and E all show some of the highest levels of deuteration in the ISM, with a number of D/H ratios $10^{5}$ times higher than Solar. We investigate for the first time the deuteration levels of the simplest COM, methanol (CH$_3$OH), in IRAS\,16293E. }
  {Using the Arizona Radio Observatory (ARO) 12\,m telescope, we target favorable transitions of CH$_2$DOH, CHD$_2$OH, $^{13}$CH$_3$OH, and several higher complexity COMs (including acetaldehyde, CH$_3$CHO, methyl formate, HCOOCH$_3$, and dimethyl ether, CH$_3$OCH$_3$) in the 3~mm band. Follow-up observations with the Yebes 40\,m telescope provided additional transitions in the 7mm (Q-band). }
  {We report the first detections of these COMs and deuterated methanol in prestellar core IRAS\,16293E and use our observations to calculate excitation temperatures, column densities, and relative abundance ratios. Striking similarities are found between relative molecular ratios and D/H values when comparing IRAS 16293E to the A and B protostars, as well as to a heterogeneous sample of other prestellar cores, protostars, and the comet 67P/Churyumov-Gerasimenko.}
  {Our results support the idea that there is a limited amount of chemical reprocessing of COMs when prestellar cores collapse and heat-up during the protostellar phase.}

               
\maketitle

%

\section{Introduction}

Cold ($\sim10$~K) and dense ($\sim 10^{5}~\mathrm{cm}^{-3}$) embryonic clumps of gas and dust, known as starless and, in the case of dynamically evolved structures (e.g., \citealt{2005ApJ...619..379C, 2008ApJ...683..238K}), prestellar cores, will collapse due to gravity and external pressure to form stars like our Sun ($M <$ a few solar masses; \citealt{2000prpl.conf...59A}). The generic term `starless' is replaced by `prestellar' if the core presents evidence that it has overcome turbulence and thermal and magnetic pressure to eventually dynamically collapse due to gravity and external cloud pressure to form an infant star. 

Interstellar complex organic molecules (COMs) contain carbon and at least 6 atoms \citep{2009ARA&A..47..427H}, and have now been detected in the gas phase toward several starless and prestellar cores across various low-mass star-forming regions \citep{2012A&A...541L..12B, 2014ApJ...795L...2V, 2016ApJ...830L...6J, 2020ApJ...891...73S, 2021MNRAS.504.5754S, 2021ApJ...917...44J, 2023MNRAS.519.1601M, 2024MNRAS.533.4104S}. COMs are thought to be key precursor species that help us understand the origins and evolution of organic chemistry, the basis for life on Earth (e.g., acetaldehyde, CH$_3$CHO, is important for the formation of alanine and other amino acids from carbonaceous chondrites; \citealt{2021PhDT........17C}). 

Understanding the degree of inheritance and/or reprocessing of COMs from the prestellar to the star- and planet-forming stages is of great interest to the astrochemistry community. Similarities in observed COM abundances when normalized to methanol, CH$_3$OH, which seems to survive the disk-formation process \citep{2016ApJ...823L..10W, 2021NatAs...5..684B, 2025ApJ...982...62E}, have been found when comparing prestellar sources to protostars as well as comets (e.g., \citealt{2019MNRAS.490...50D, 2021MNRAS.504.5754S, 2024MNRAS.533.4104S}). These similarities suggest that when stars and planets continue forming and evolving within their original envelope of molecular gas and dust, at least part of their chemical composition is determined by the inheritance of molecules from the preceding evolutionary stages. Thus, to understand the astrochemical origins of the composition of planetary systems, we must set constraints on the chemical evolution at the onset of cloud and star formation, i.e., in cold starless and prestellar cores (see \citealt{2021PhR...893....1O} and references therein). 

Deuterated COMs (D-COMs), i.e., COMs that contain deuterium, provide an exciting probe into the chemical history, as high deuteration levels in protostellar systems (e.g., IRAS~16293-2422\,A and B with D/H~$\sim$ 2-8\% and D$_2$/D~$\sim$ 20\%; \citealt{2018A&A...620A.170J, 2022A&A...659A..69D}), compared to the elemental value in the interstellar medium (ISM; D/H $\sim 10^{-5}$; \citealt{2006ApJ...647.1106L}), suggest that COMs form during a time of enhanced deuteration. It is in the cold ($\sim$~10\,K) prestellar core stage that deuterium fractionation is predominately efficient, as deuterated molecules have lower zero point energies compared to molecules with hydrogen \citep{2002P&SS...50.1257C, 2007A&A...470..221C}. At these lower temperatures CO (the main destroyer of H$_{3}^+$ and H$_2$D$^+$; \citealt{1989MNRAS.237..661B, 2003ApJ...591L..41R, 2003A&A...403L..37C}) will also freeze out or `deplete' onto the grains allowing for this enhancement of deuterated molecules \citep{2002P&SS...50.1133C, 2005ApJ...619..379C}, including larger species such as doubly-deuterated formaldehyde, D$_2$CO \citep{2003ApJ...585L..55B}. 
However, to date, only a handful of prestellar cores have had detections of both singly- and doubly-deuterated versions of the simplest COM methanol (e.g., H-MM1 \& L694-2 with {CH$_2$DOH/CH$_3$OH} $\sim$ 0.8-1.9\% and {CHD$_2$OH/CH$_2$DOH} $\sim$ 50-80\%; \citealt{2023A&A...669L...6L}).

The prestellar core IRAS\,16293E, located within the L1689N molecular cloud in the Rho Ophiuchi cloud at a distance of $\sim$141\,pc \citep{2018A&A...614A..20D}, is believed to have the highest deuteration levels among starless and prestellar cores, based on strong emission of triply-deuterated ammonia, ND$_3$ \citep{2016ApJ...827..133L}. Earlier observations also found large amounts of D$_2$CO toward this core \citep{2001ApJ...552L.163L}, as well as toward several positions in the molecular cloud L1689N itself \citep{2002A&A...381L..17C}. Other deuterated species such as DCO$^+$ \citep{1987ApJ...317..220W}, DCN \citep{2023A&A...673A.143K}, NH$_2$D \citep{2001ApJ...554..933S}, ND$_2$H \citep{2001ApJ...552L.163L, 2006ApJ...636..916L, 2006A&A...454L..63G}, H$_2$D$^{+}$ \citep{2004ApJ...608..341S, 2008A&A...492..703C, 2014Natur.516..219B}, D$_2$H$^{+}$ \citep{2004ApJ...606L.127V, 2024A&A...691A..88P}, N$_2$D${+}$ \citep{2024arXiv241213760S}, and HDO \citep{2004ApJ...608..341S} are also prevalent in the core and/or surrounding cloud. 

IRAS\,16293E sits $90\arcsec$ ($12~700$~au) to the east of its protostellar neighbors, IRAS\,16293-2422 A and B \citep{2004ApJ...608..341S}. Early observations of the protostellar system revealed high abundances of singly-deuterated methanol, CH$_2$DOH, as well as provided the first ISM detections of doubly-deuterated methanol, CHD$_2$OH \citep{2002A&A...393L..49P} and triply-deuterated methanol, CD$_3$OH \citep{2004A&A...416..159P}. More recently, the ALMA Protostellar Interferometric Line Survey (PILS) team have uncovered a rich inventory of even larger gas phase COMs and associated D-COMs, including the simple sugar glycolaldehyde \citep{2012ApJ...757L...4J, 2016A&A...595A.117J}. 

Observations of simpler molecular gas tracers (e.g., CH$_3$OH, SO, CS, and H$^{13}$CO$^+$) around both the IRAS\,16923-2422 protostars and IRAS\,16293E show clear differences in spatial distribution, derived column densities, and temperatures, revealing the diverse physical conditions in the cloud complex \citep{2023A&A...673A.143K}. 
Furthermore, it is thought that the large-scale outflows that are driven by IRAS\,16293-2422\,A are interacting with IRAS\,16293E. Due to the impinging outflows, gas phase species may more readily be desorbed off the grains. Gas phase methanol, CH$_3$OH, has already been robustly detected toward IRAS\,16923E with APEX observations (277--375\,GHz) of higher upper energy (E$_u \sim 17-50$\,K) transitions \citep{2023A&A...673A.143K, 2024arXiv241213760S}, yet there have not been observations dedicated to detecting any other larger COMs or D-COMs.

IRAS\,16293E is a unique core whose ice chemistry and gas phase chemistry are, both, being thoroughly investigated. The executed JWST Cycle 2 General Observer (GO) program \#3222 "Cask-strength clouds: high percentage of methanol and HDO ices" (P.I.: M.~N. Drozdovskaya) will probe the chemical composition of ices in the L1689N dark cloud that houses the prestellar core IRAS\,16293E (Drozdovskaya et al. in prep.). 
Here, we present an Arizona Radio Observatory (ARO) 12\,m and Yebes 40\,m COM emission line survey toward IRAS\,16923\,E and focus this study on reporting column density estimates for newly detected singly-deuterated methanol (CH$_2$DOH), doubly-deuterated methanol (CHD$_2$OH), the $^{13}$C-isotopologue of methanol ($^{13}$CH$_3$OH), as well as the COMs acetaldehyde (CH$_3$CHO), methyl formate (HCOOCH$_3$), and dimethyl ether (CH$_3$OCH$_3$). We also derive the column density of methanol from low-energy ($E_\mathrm{u}$<20\,K) transitions.
In Section\,\ref{sec:obs}, we describe the observations and data reduction methods. In Section\,\ref{sec:results}, we calculate column densities and excitation temperatures. We then put our calculations into context with other sources in Section\,\ref{sec:discuss} and conclude in Section\,\ref{sec:conclude}.

\begin{figure}
   \centering
    \includegraphics[width=90mm]{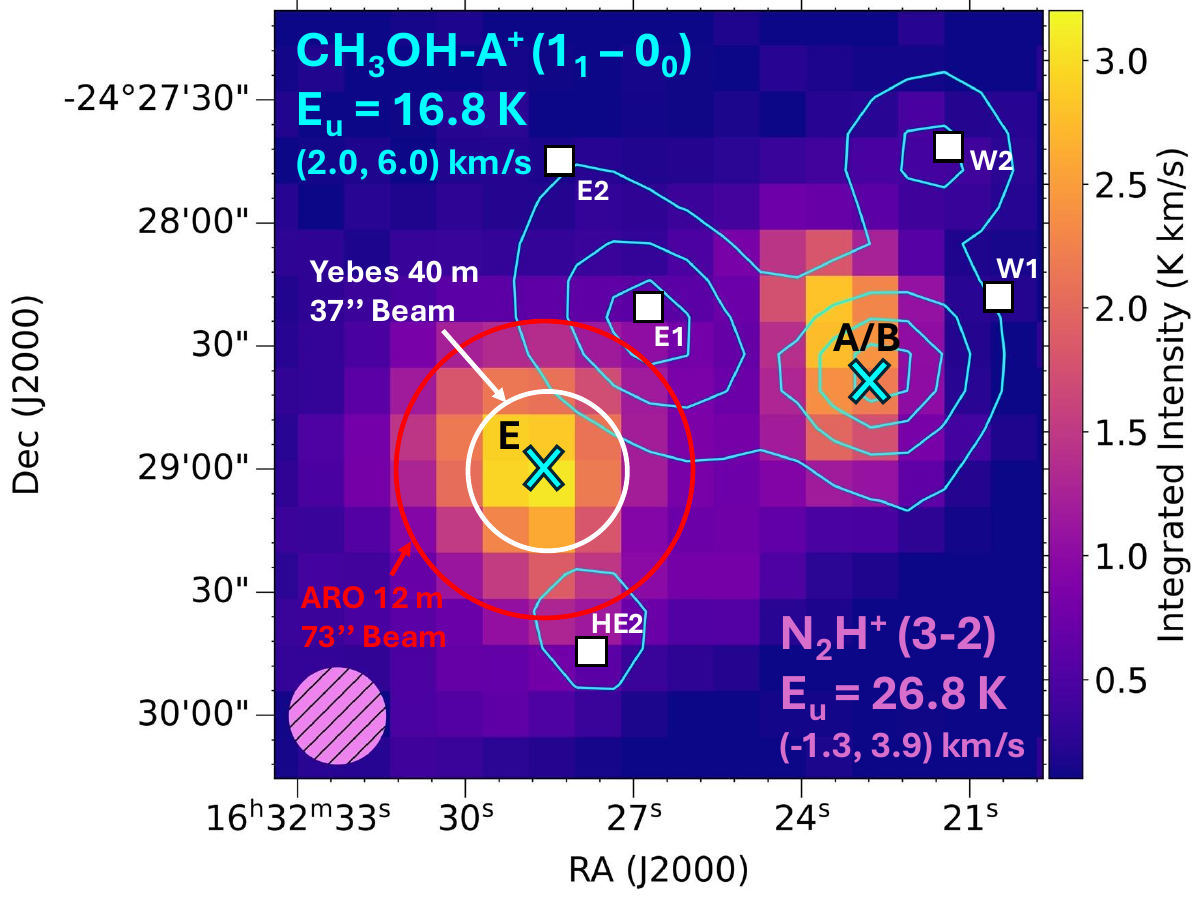} 
    \caption{Map of the IRAS 16293-2422 region from publicly available APEX data presented in \citealt{2023A&A...673A.143K}, where the colormap is N$_2$H$^{+}$ (3-2) at 279.511\,GHz and the cyan contours show the CH$_3$OH-A$^{+}$ $1_1 - 0_0$ b-type transition at 350.905\,GHz (beam size for N$_2$H$^{+}$ is 23.7$^{''}$ and shown in bottom left). Crosses designate where the IRAS 16293E prestellar core, labeled `E', and the IRAS 16293A and B protostars labeled as `A/B', sit. Other emission regions studied in \citealt{2023A&A...673A.143K} are labeled by white squares. For the single-pointing observations presented here, we also show the largest beam size (ARO 12 m beam at 73$^{''}$) and the smallest  beam size (Yebes 40 m beam at 37$^{''}$). }
    \label{fig:map}
\end{figure}

\begin{table*}[h!] 
\caption{ COM lines targeted with ARO 12 m and selection of COM lines targeted with Yebes 40\,m }            
\label{tab:lines}     
\centering                         
\setlength{\tabcolsep}{5pt}
\begin{tabular}{l l l l l l l l l l l l}       
\hline\hline         
Molecule & Transition & Frequency & $E_\mathrm{u}$ & $A_{\mathrm{ul}}$ & $g_\mathrm{u}$ & $T_\mathrm{mb}$   & \textit{rms} & Vel & FWHM  & $\theta_\mathrm{beam}$\\ [1pt] 
 &   &  (GHz) &  (K) & (s$^{-1}$) &   & (mK) & (mK) & (km s$^{-1}$) & (km s$^{-1}$) & (${''}$)\\ [3pt] 
\hline                 
\multicolumn{10}{c}{\textbf{ARO 12\,m}} \\ 
\hline 
CH$_3$OH &  2$_{-1,2} - 1_{-1,1}$  E& 96.739358(2) & 12.5 & 2.6E-06 & 20 & 1011.9 & 4.9 & 3.76{($<$1)} & 1.10{($<$1)} & 62.3 \\
     &  2$_{0,2} - 1_{0,1}$ A$^+$ & 96.741371(2) & 6.7 &  3.4E-06  & 20 & 1223.7 & 4.9 & 3.71{($<$1)} & 1.12{($<$1)} & 62.3 \\
   &  2$_{0,2} - 1_{0,1}$ E & 96.744545(2)  & 20.1  & 3.4E-06  & 20 & 206.7 & 4.9 & 3.75{($<$1)} & 1.20{($<$1)} & 62.3 \\ 

$^{13}$CH$_3$OH & $2_{-1, 2} - 1_{-1,1}$ E & 94.40516(5) & 12.4& 2.3E-06 & 5 & 19.7 & 5.9 &  3.49(1) & 0.96(3) & 63.9 \\ 
                & $2_{0, 2} - 1_{0,1}$ A & 94.407129(4) & 6.7 & 3.1E-06 & 5 & 30.6 & 5.0 &  3.59(1) & 0.74(2) & 63.9 \\  
                
CH$_2$DOH  &  $1_{1,0} - 1_{0,1}$ e$_0$  &  85.29690(10) & 6.2 & 4.5E-06& 3 & 40.8 &4.5 & 3.77{($<$1)} & 0.57(1) & 70.9 \\
& $2_{1,1} - 2_{0,2}$ e$_0$   &  86.66886(10) & 10.6 & 4.6E-06& 5 & 56.0 & 5.2 &  3.81{($<$1)} & 0.90(1) & 69.6 \\
        &  $2_{1,2} - 1_{1,1}$ e$_0$      &  88.07312(10) & 10.4 & 1.4E-06 & 5 & 15.6 & 3.8 & 3.64(1) & 0.52(2) & 68.5\\ 
&  $2_{0,2} - 1_{0,1}$ e$_0$   &  89.40791(16) & 6.4 & 2.0E-06& 5 & 64.0 & 4.6 & 3.63{($<$1)} & 0.62{($<$1)} & 67.4\\ 

CHD$_2$OH  &    $2_{1,2} - 1_{1,2}$ e$_0$   & 82.16582(10) & 9.1 & 1.6E-06 & 5 & 16.7 & 5.8 & 3.77(1) & 0.36(2) & 73.4 \\
           &    $2_{0,1} - 1_{0,1}$ e$_0$   & 83.28963(10) & 6.0 & 2.2E-06 & 5 & 24.8 & 4.1 & 3.54(1) & 0.34(1) & 72.4 \\  

 CH$_3$CHO & 2$_{1,2} - 1_{0,1}$ A & 84.219749(3)  & 5.0 &  2.4E-06 & 10 & -- & 4.7 & -- & -- & 71.6 \\
 & 5$_{1, 5} - 4_{1, 4}$ A & 93.580909(3) & 15.7 & 2.6E-05 & 22& 42.0 & 6.2 & 3.72(1) & 0.95(1)  & 64.4 \\
  & 5$_{1, 5} - 4_{1, 4}$ E & 93.595235(3) & 15.8 & 2.6E-05 &22 & 48.5 & 6.9 & 3.80{($<$1)} & 0.86(1)& 64.4  \\ 
         &  5$_{0,5} - 4_{0,4}$ E  & 95.947437(3)  & 13.9  & 3.0E-05 &22 & 51.7 & 5.0 & 3.81{($<$1)} & 0.90(1) & 62.8 \\
    & 5$_{0,5} - 4_{0,4}$ A  &  95.963459(3) & 13.8 & 3.0E-05 &  22 & 54.0 & 5.3 &3.85{($<$1)} & 0.98(1) & 62.8 \\ 
HCOOCH$_3$ &   $8_{1,7} - 7_{1,6}$ A\tablefootmark{a}  & 100.49068(1) & 22.7 & 1.4E-05 & 34 & 10.5& 3.3 & 4.33(6) & 2.73(15) & 60.0 \\ 
           &   $8_{1,7} - 7_{1,6}$ E\tablefootmark{a}  & 100.48224(1) & 22.7 & 1.4E-05 & 34  & 12.0 & 3.8 & 3.66(3) & 0.97(8) & 60.0   \\ 
           
CH$_3$OCH$_3$ 
&  4$_{1,4} - 3_{0,3}$ AA & 99.32607(10) & 10.2 & 5.5E-06 & 90 & -- & 5.7 & --& -- & 60.7 \\ 
              &  4$_{1,4} - 3_{0,3}$ EE  & 99.32522(10) & 10.2 & 5.5E-06 & 144 & -- & 5.7 & -- & -- & 60.7 \\ 
              &  4$_{1,4} - 3_{0,3}$ AE+EA & 99.32436(10) & 10.2 & 5.5E-06 & 90  & -- & 5.7 & --& --  & 60.7 \\  

CH$_2$CHCN  & 9$_{1,8} - 8_{1,7}$ & 87.312807(30) & 23.1 & 5.3E-05 & 57 & -- &  4.3 & -- & -- & 69.0\\ 
&$10_{0,10} - 9_{0,9}$&	94.276624(40) & 24.9 &  6.2E-05	& 63 & -- & 4.1 & -- & -- & 64.0 \\
&	$10_{1,9} - 9_{1,8}$&	96.982447(40) & 27.8 &  7.2E-05	& 63 & -- & 7.7 & -- & -- & 62.1 \\ 
\hline                
\multicolumn{10}{c}{\textbf{Yebes 40\,m}} \\   
\hline 
 CH$_3$OH   &  1$_{0,1} - 0_{0,0}$  A & 48.372460(1)  & 2.3 &  3.6E-07 & 12  & 778.3 & 7.3 & 3.75{($<$1)} &  0.96{($<$1)} & 37.4 \\
     &  1$_{-0,1} - 0_{-0,0}$  E &  48.376887(1) & 15.4 &   3.6E-07  &12 & 116.2  & 9.3 & 3.68{($<$1)} & 0.93{($<$1)} & 37.4 \\  
     
CH$_2$DOH  & 1$_{0,1} - 0_{0,0}$ e$_0$ & 44.713190(80) &  2.1 &  2.1E-07 & 3 & 26.5 & 5.2 &  3.33(1) & 0.69(3) & 40.5 \\
 CH$_3$CHO & 2$_{1,2} - 1_{1,1}$ A  & 37.464204(1)  & 4.9  & 1.2E-06 & 10 & 32.3 & 2.0 & 3.86(1) & 0.88(3) & 48.3 \\
&         2$_{1,2} - 1_{1,1}$ E  & 37.686932(2) & 5.0  & 1.1E-06 & 10 &  19.2 & 2.5 & 3.80(2) &  0.78(4) & 48.0 \\
&        2$_{0,2} - 1_{0,1}$ E & 38.506034(1)  &  2.8  & 1.7E-06 & 10 & 52.2 & 2.1 & 3.88(1) & 0.84(2) & 47.0 \\
&       2$_{0,2} - 1_{0,1}$ A & 38.512079(1) &  2.8 & 1.7E-06 &  10 & 54.3 & 2.6 & 3.91(1) & 0.73(1) & 47.0 \\
&      2$_{1,1} - 1_{1,0}$ E & 39.362537(2)  & 5.2  &  1.3E-06  & 10 & 31.6 & 2.0 & 3.79(1) & 0.82(3) & 46.0 \\
&      2$_{1,1} - 1_{1,0}$ A & 39.594289(1)  & 5.1  & 1.4E-06& 10 &  35.6 & 3.1 & 3.90(1) &  0.82(2) & 46.0 \\
HCOOCH$_3$  & 3$_{0, 3}- 2_{0, 2}$ E & 36.102224(50) & 3.5 &  6.2E-07 &  14 & 6.9 & 1.9 & 3.81(5) & 0.76(11) & 50.1 \\
&     3$_{0, 3}- 2_{0, 2}$ A & 36.104793(50) & 3.5 & 6.2E-07 & 14 &  8.9 & 2.9 & 3.88(4) &  0.91(1) & 50.1 \\ 

CH$_3$OCH$_3$ &  $3_{1, 2} - 3_{0, 3}$ AE+EA & 32.977274(1)  & 7.0 & 3.4E-07 & 70  & 6.8 & 1.7 & 3.60(6) & 1.04(14) & 54.8 \\  
& $3_{1, 2} - 3_{0, 3}$ EE & 32.978232(1) & 7.0 & 3.4E-07 & 112  & 8.6 & 1.5 & 3.78(4) & 0.67(8) & 54.8 \\
& $3_{1, 2} - 3_{0, 3}$ AA & 32.979187(1)  & 7.0 & 3.4E-07 & 70  & 6.1 & 1.6 & 4.21(5) & 0.56(13) & 54.8 \\ 

& $5_{1, 4} - 5_{ 0, 5}$ AE+EA & 39.046259(1)  & 15.4 & 5.0E-07 & 110 & -- & -- & -- & -- & 46.3 \\
& $5_{1, 4} - 5_{ 0, 5}$ EE &  39.047303(1) & 15.4  &	5.0E-07 & 176 &  10.6 & 1.9 &  4.19(3) & 0.70(7) & 46.3 \\
& $5_{1, 4} - 5_{ 0, 5}$ AA & 39.048346(1) &  15.4  &	5.0E-07 & 110 & -- & -- & -- & -- & 46.3 \\
\hline                                   
\end{tabular}
\tablefoot{ Line information {including frequency, upper energy ($E_\mathrm{u}$), spontaneous emission coefficient ($A_\mathrm{ul}$) and the upper state degeneracy ($g_\mathrm{u}$) }for CH$_3$CHO and HCOOCH$_3$ is from the JPL catalog 
     (\url{https://spec.jpl.nasa.gov/}; \citealt{1998JQSRT..60..883P}) and for the remaining transitions from the CDMS database (\url{https://cdms.astro.uni-koeln.de};  \citealt{2001A&A...370L..49M, 2005JMoSt.742..215M, 2016JMoSp.327...95E}). The laboratory spectroscopic work for CH$_3$OH is in \cite{2008JMoSp.251..305X}, for $^{13}$CH$_3$OH in \cite{1997JPCRD..26...17X}, for CH$_2$DOH in \cite{2012JMoSp.280..119P, 2014JChPh.140f4307C}, for CHD$_2$OH in \cite{2022A&A...659A..69D}, for CH$_3$CHO in \cite{1996JPCRD..25.1113K}, HCOOCH$_3$ in \cite{1979JPCRD...8..583B, 2001JMoSp.210..196K, 2004JMoSp.225...14O, 2009JMoSp.255...32I}, CH$_3$OCH$_3$ in \cite{1976JMoSp..62..159D, 1979JPCRD...8.1051L, 2009A&A...504..635E}, and for CH$_2$CHCN in \cite{1985ZNatA..40..998S, 1988JMoSp.130..303C}. 
     {The Gaussian fitting parameters for IRAS\,16293E are also listed next to each line transition, where ‘$T_\mathrm{mb}$’ represents the peak intensity on the main temperature beam scale, ‘rms’ is the 1$\sigma$ standard deviation in the noise level, ‘Vel’ is the line-of-sight centroid velocity, and ‘FWHM’ is the full width at half maximum {of the line}. The telescope beam, $\theta_\mathrm{beam}$, is listed in the last column.}
     Numbers in parentheses denote 1$\sigma$ uncertainties in unit of the last quoted digit, where a `$<$1' indicates an uncertainty in the fitting routine (with Pyspeckit) of less than 0.01\,km/s for the centroid velocity or FWHM. \tablefoottext{a}{Smoothed by 3 channels.}} 
\end{table*}

\begin{figure} 
\begin{center}$
\centering
\begin{array}{c}
\includegraphics[width=85mm]{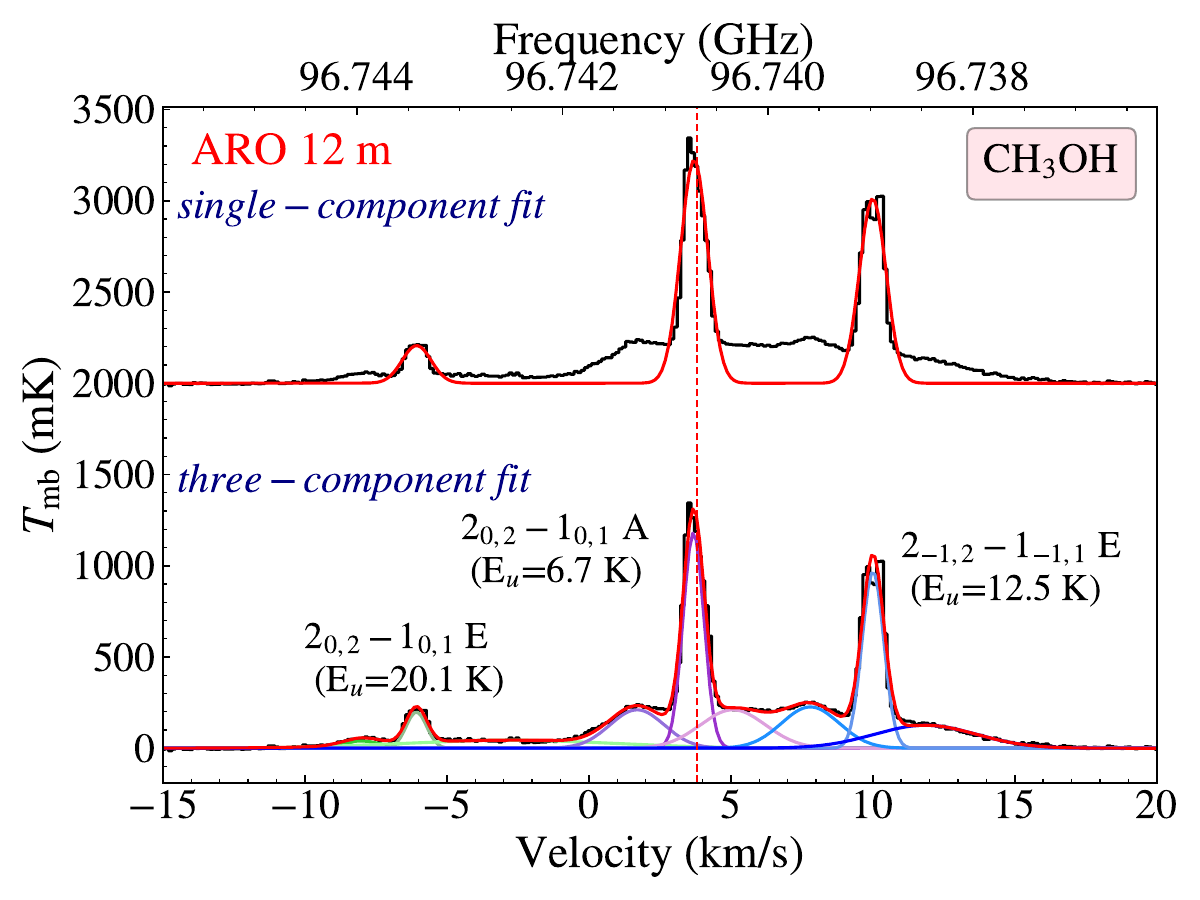} 
\end{array}$
\end{center}
    \caption{Observed CH$_3$OH spectrum toward IRAS 16293E shown in black and Gaussian fit as the red curve. A vertical red line at v$_{lsr}$ of 3.8~km/s is centered on the strongest line with lowest $E_{u}$. (top) A single-component fit, as reported in Table\,\ref{tab:lines} and (bottom) a three-velocity fit to each transition, separated out into the green, purple and blue curves with a total red composite curve, which is better able to reproduce the full profile; however we note that the centrally peaked profile (i.e., at the v$_{lsr}$ of 3.8\,km/s) from our single component fit remains the most dominant.} \label{fig:ch3oh}
\end{figure}

\section{Observations} \label{sec:obs}

We detail below our single-pointing observations with both the ARO 12 m and Yebes 40 m telescopes that span beam sizes from roughly $37^{''}$ to $73^{''}$ ($5~200$ to $10~3000$~au) toward the IRAS 16293E prestellar core in the L1689N molecular cloud  (see Figure\,\ref{fig:map} for scale). 

\subsection{ARO 12 m} \label{sec:ARO12m}

Molecular line observations were obtained with the ARO 12\,m radio dish on Kitt Peak outside of Tucson, Arizona. Data was collected from November 2023 to January 2024 with the $3$~mm sideband separating dual polarization receiver. Each scan was 5 minutes using the standard absolute position switching (APS) observing mode between our source, i.e., the N$_2$H$^+$ molecular emission peak position of IRAS\,16293E (RA, DEC (J2000): $16^\mathrm{h}32^\mathrm{m}28.5^\mathrm{s}$, $-24^\mathrm{o}29\arcmin02.0\arcsec$; {\citealt{2001A&A...375...40C, 2023A&A...673A.143K}}), and the off position void of emission (RA: $16^\mathrm{h}32^\mathrm{m}33^\mathrm{s}$, DEC: $-24^\mathrm{o}25\arcmin00.0\arcsec$) every 30 seconds. Pointing was checked every $\sim$1-2 hours on the nearby planet Venus. 

The AROWS spectrometer, with a spectral resolution of 39\,kHz (0.12 km/s -- 0.14 km/s) and an automatic Hanning smoothing applied, was used for all observations with the two polarizations (vertical and horizontal). In setups 1 (Local Oscillator, LO, at 90\,GHz) and 2 (LO at 93.5\,GHz), we were able to observe eight lines in parallel, using the multi-window mode where each line was placed in a separate 250\,MHz bandwidth spectral window, with groups of four within 4\,GHz either in the Lower Side Band (LSB) or in the Upper Side Band (USB). 
In setup 3 (LO at 101\,GHz), we were not able to use one of the AROWS boards and only four lines could be tuned simultaneously in the LSB. 
See Table\,\ref{tab:lines} for the parameters of the covered lines. 

Initial reduction (i.e., combining of spectral scans) was performed using the CLASS program of the GILDAS package (\citealt{2005sf2a.conf..721P},\citealt{2013ascl.soft05010G})\,\footnote{\url{http://iram.fr/IRAMFR/GILDAS/}}. For each of the two polarizations, a median beam efficiency percentage was calculated, along with estimated errors from median measurements of Venus. Both polarization measurements produced equivalent efficiencies, and thus a single $\eta_\mathrm{ARO} =  92.15 \pm 2.74\%$ was adopted. The spectra were then converted to the main-beam temperature using the equation $T_{mb}$ = $T_A^*$/$\eta_\mathrm{ARO}$ \citep{1993PASP..105..117M}. This scaling and Gaussian fitting of spectral lines was done with the python `Pyspeckit' package \citep{2011ascl.soft09001G, 2022AJ....163..291G}. 

\begin{figure*}
\begin{center}$
\begin{array}{c}
\includegraphics[width=180mm]{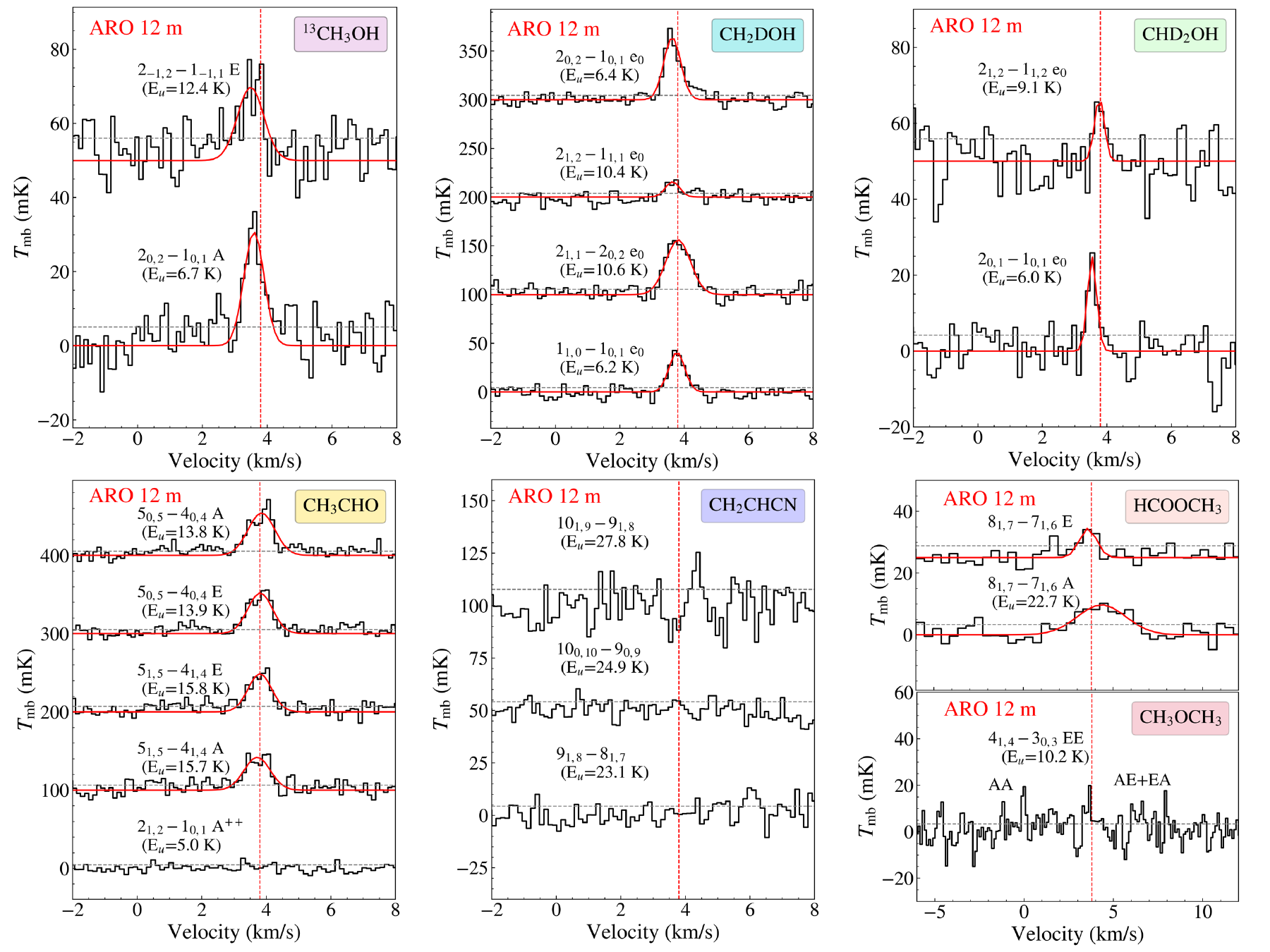}
\end{array}$
\end{center}
\caption{Observed ARO 12\,m spectra toward IRAS 16293E shown in black and Gaussian fits as red curves. Spectrum with no Gaussian fits are considered non-detections and upper limits are derived. Vertical red lines are centered at v$_{lsr}$ of 3.8 km/s and the gray horizontal lines show the $1\sigma$ noise ($rms$) level. From top left across to bottom right: $^{13}$CH$_3$OH spectra offset by 50\,mK, CH$_2$DOH spectra offset by 100\,mK, CHD$_2$OH spectra offset by 50\,mK, CH$_3$CHO spectra offset by 100\,mK, CH$_2$CHCN spectra (non-detections) offset by 50\,mK, HCOOCH$_3$ spectra offset by {25}\,mK, and a single CH$_3$OCH$_3$ spectrum (non-detection). }
\label{fig:12m_total}
\end{figure*}

For simplicity, a single velocity component is assumed in the Gaussian fitting for each line of interest, even for CH$_3$OH and CH$_3$CHO which clearly show multiple velocity components (Figures\,\ref{fig:ch3oh}\,and\,\ref{fig:12m_total}), likely because these species are more widespread and our large beam ($\sim$60$^{''}$) picks up gas affected by interactions with the nearby protostellar outflows coming from the direction of the A/B protostars (Figure\,\ref{fig:map}). 
Additionally, for HCOOCH$_3$ we degraded the spectral resolution of the spectra by a factor of 3 to increase the signal-to-noise ratio for these weak lines, leading to more uncertainty in linewidth and centroid velocity (Figure\,\ref{fig:12m_total}).
All line fitting results are listed in Table\,\ref{tab:lines}, where the noise level or \textit{rms} is calculated by finding the standard deviation in the spectrum away from any line emission.  

\subsection{Yebes 40 m} \label{sec:Yebes40m}

Data from the dual (horizontal and vertical) linear polarization Q-band receiver \citep{2021A&A...645A..37T} on the Yebes 40\,m telescope was taken during the spring of 2024 (24A006: PI Scibelli). These observations were done in position switching mode with the same off position that was used during the ARO 12\,m observations. We note that while frequency-switching is also available, confusion from baseline ripples was a concern, as certain lines could be broader and confused due to the nearby outflow and multiple velocity components. The wide-band nature of the receiver allows for a total bandwidth of 18.5 GHz spanning 31.5 -- 50 GHz (6 -- 9~mm) with a resolution of 38.0~kHz (0.38~km/s -- 0.23~km/s). 

The data were inspected, reduced, and put on the main beam temperature, $T_\mathrm{mb}$, scale using publicly available Python-based scripts\footnote{\url{https://github.com/andresmegias/gildas-class-python/}} developed by \cite{2023MNRAS.519.1601M}, which invokes the CLASS program of the GILDAS package for the combining of the spectra. The main beam efficiencies measured and given by Yebes are 0.65(0.64) at 32.4 GHz, 0.61(0.61) at 34.5 GHz, 0.61(0.60) at 36.9 GHz, 0.59(0.58) at 39.2 GHz, 0.58(0.57) at 41.4 GHz, 0.56(0.55) at 43.7 GHz, 0.54(0.54) at 46.0 GHz and 0.52(0.41) at 48.4 GHz for the horizontal(vertical) polarizations.
As for the ARO 12\,m spectra, the Gaussian fitting of spectral lines was done with the python `Pyspeckit' package \citep{2011ascl.soft09001G, 2022AJ....163..291G} and the fitting results are listed in Table\,\ref{tab:lines}. 
In general, the $\sigma_{T_\mathrm{mb}}$ values range over $\sim 2 - 9$\,mK. 

\begin{figure*}
\begin{center}$
\begin{array}{c}
\includegraphics[width=180mm]{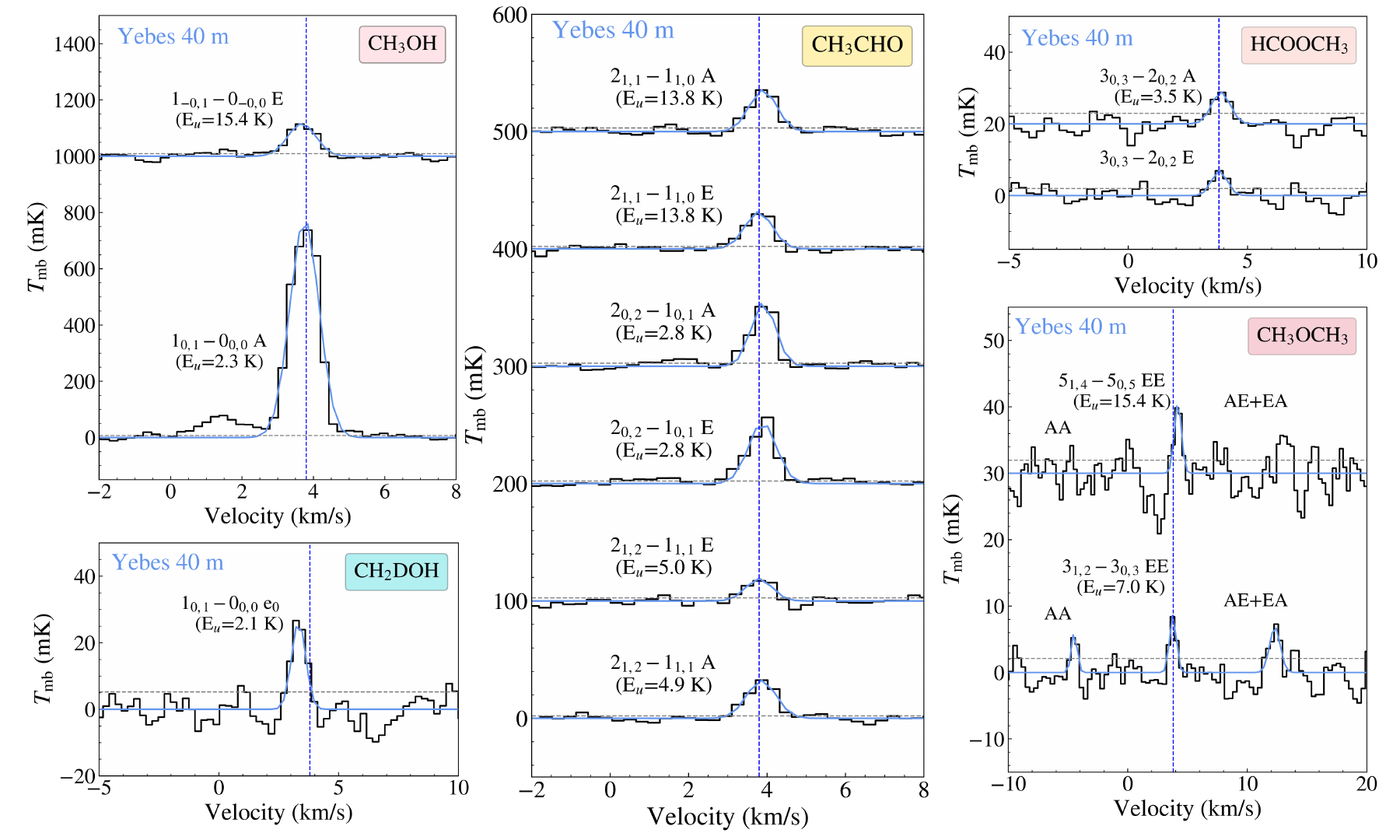}
\end{array}$
\end{center}
\caption{Observed Yebes 40\,m spectra toward IRAS 16293E shown in black and Gaussian fits as blue curves. Vertical blue lines are centered at v$_{lsr}$ of 3.8 km/s and the gray horizontal lines show the $1\sigma$ noise ($rms$) level. 
From top left down to bottom right: CH$_3$OH spectra offset by 1000\,mK, CH$_2$DOH spectrum, CH$_3$CHO spectra offset by 100\,mK, HCOOCH$_3$ spectra offset by 20\,mK, and CH$_3$OCH$_3$ spectra offset by 30\,mK.
}
\label{fig:yebes_total}
\end{figure*}

For this analysis, we concern ourselves only with `energetically favorable' transitions (i.e., transitions {with} $E_{u} <$ 25\,K and Einstein $A_{ul}$ values $> 1.0 \times 10^{-7}$\,s$^{-1}$) of the COMs and D-COMs already targeted with the ARO 12\,m, which include CH$_3$OH, CH$_2$DOH, CH$_3$CHO, HCOOCH$_3$, CH$_3$OCH$_3$, and CH$_2$CHCN (which remains undetected). 
These Yebes Q-band COM transitions (as well as a single ARO 12m setup that includes the CH$_3$OH 96.7\,GHz, CH$_3$CHO 95.9\,GHz, and CH$_2$CHCN 94-96\,GHz transitions) were previously investigated toward starless and prestellar cores in the Perseus Molecular Cloud \citep{2024MNRAS.533.4104S}. 
While other lines {are covered in} our band, they will be presented in a subsequent paper.

\section{Column densities}\label{sec:results}

We report the first detection of the methanol isotopologues $^{13}$CH$_3$OH, CH$_2$DOH, and CHD$_2$OH, as well as the COMs CH$_3$CHO, HCOOCH$_3$, and CH$_3$OCH$_3$ toward the emission peak of IRAS\,16293E via multiple transitions from both the ARO 12\,m (Figure\,\ref{fig:12m_total}) and the Yebes 40\,m (Figure\,\ref{fig:yebes_total}). We also report non-detections and an upper limit for CH$_2$CHCN (Figure\,\ref{fig:12m_total}). Below we detail our column density calculations. 

\begin{figure}
\begin{center}$
\begin{array}{ccc}
\includegraphics[width=80mm]{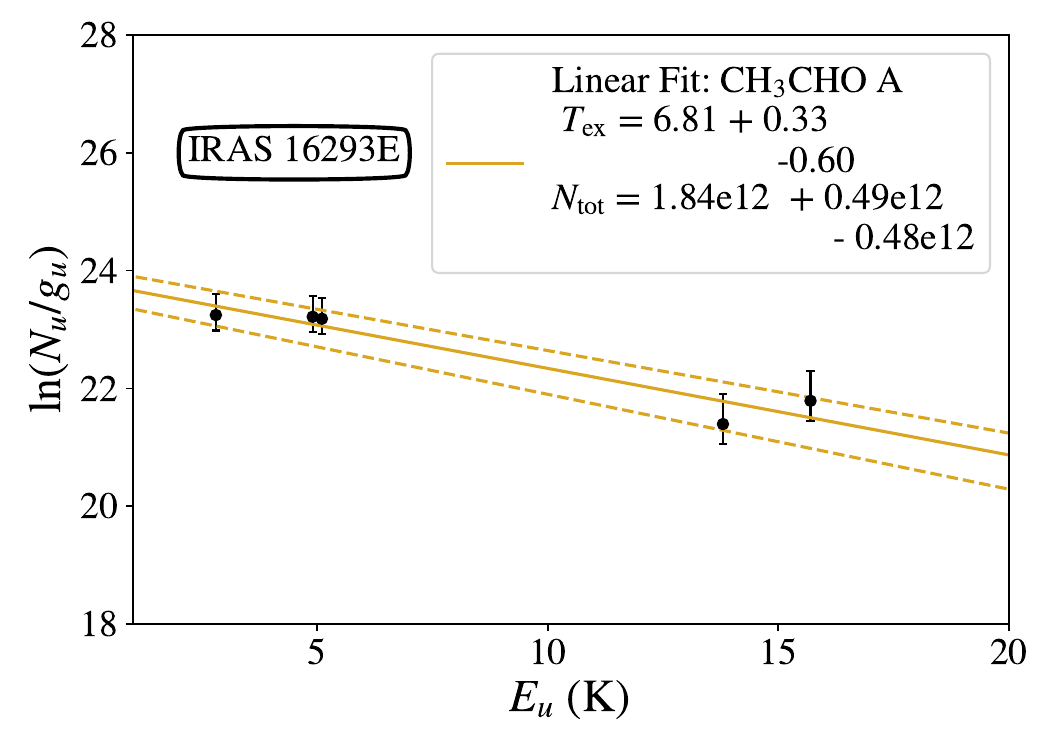}\\ 
\includegraphics[width=80mm]{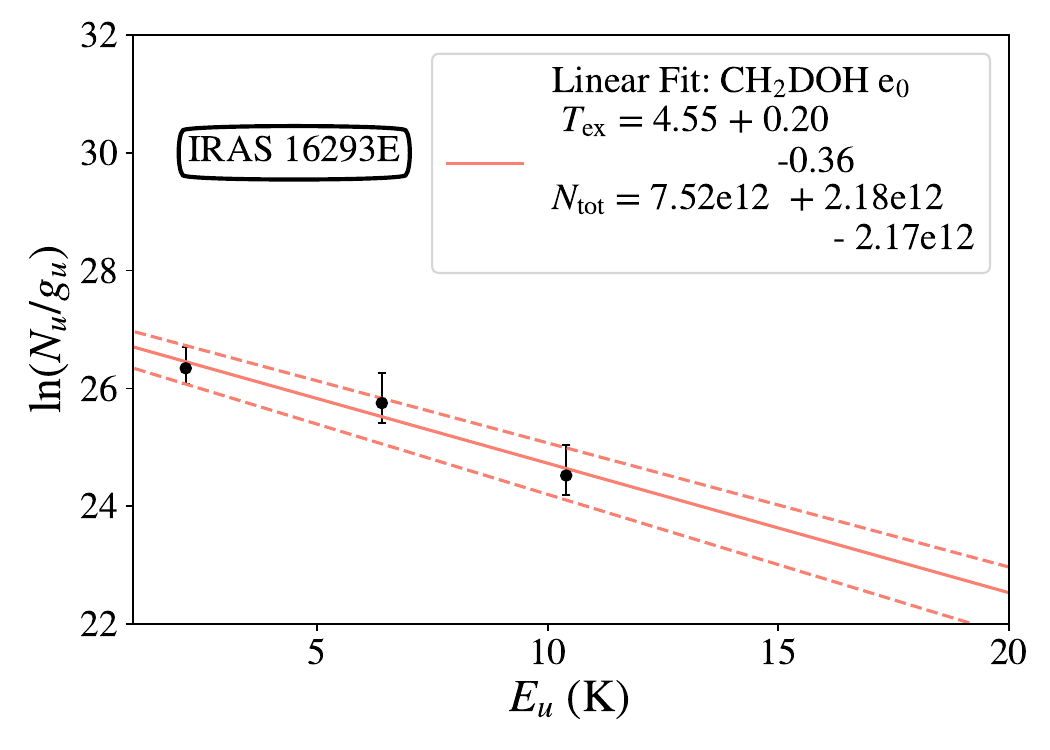}
\end{array}$
\end{center}
\caption{ Rotation diagrams with associated linear best-fits (solid curves) and corresponding uncertainty (dashed curves) for (top) CH$_3$CHO A that utilizes five transitions and for (bottom) CH$_2$DOH that utilizes three transitions.
}
\label{fig:rotatation}
\end{figure}

\subsection{{RADEX method: CH$_3$OH, $^{13}$CH$_3$OH, and HCOOCH$_3$}} \label{sec:subRADEX}

We begin by using the non-Local Thermodynamic Equilibrium (non-LTE) RADEX code that assumes a uniform isothermal cloud distribution \citep{2007A&A...468..627V} in order to calculate column densities, $N$, and excitation temperatures, $T_\mathrm{ex}$, while minimizing over the input kinetic temperature of the gas, $T_\mathrm{k}$, and the source size, $\theta_\mathrm{src}$, of the emitting area when possible (i.e., when more than a single transition is available). This $\theta_\mathrm{src}$ relates to the fact that the molecular emission from our source likely does not uniformly fill the telescope beam, $\theta_\mathrm{beam}$. This is accounted for by dividing the observed integrated intensity by a filling factor, \textit{f}, which measures the coupling between the source integrated intensity emission, $I(\theta, \phi)$, and the telescope beam, $P_n(\theta)$. As described in \cite{2013tra..book.....W}, by assuming the source emission can be approximated by a Gaussian with FWHM $\theta_\mathrm{src}$, then the filling factor  is given by: \begin{equation}\label{fillingfrac}
    f = \frac{\int I(\theta,\phi) P_n(\theta) \mathrm{d}\Omega}{I(\theta=0) \int P_n (\theta) \mathrm{d}\Omega } = \frac{\theta_\mathrm{src}^2}{\theta_\mathrm{src}^2 + \theta_\mathrm{beam}^2}. 
\end{equation}

Due to the limited number of transitions for each molecule of interest (that ranged from one to three), we decided to fix the input volume density within the observed beam, $n_{\mathrm{beam}}$, in our RADEX calculations. 
We first calculate the median molecular hydrogen column density, $N_\mathrm{H_2}$, directly from the Ophiuchus \textit{Herschel} column density maps \citep{2010A&A...518L.102A, 2020A&A...638A..74L} within the given telescope beam ($\theta_\mathrm{beam}$) for each transition of interest. We find $n_{\mathrm{beam}}$ with,

\begin{equation}
    {n_\mathrm{beam} = \frac{N_\mathrm{H_2}}{l}, }
\end{equation} where $l = \theta_\mathrm{beam}d$ is the length of the beam given a distance $d$ to the source. For $\theta_\mathrm{beam}$ values that range from 37.4$^{''}$ to 63.9$^{''}$, a range of mean volume densities is calculated to be $n_{\mathrm{beam}} = (0.71 - 1.38) \times 10^6$\,cm$^{-3}$. These values are consistent with the volume density estimate of $n_{\mathrm{H}_2} = 1 \times 10^6$\,cm$^{-3}$ reported in the literature \citep{2004ApJ...608..341S}. A more recent analysis from \cite{2024arXiv241213760S} found a density $n_{\mathrm{H}_2} \sim 5 \times 10^5$\,cm$^{-3}$ at $\sim4500$\,au at a radius of $\sim4500$\,au, which corresponds to the largest angular scale that we probe (63.9\arcsec or $\sim$9,000 au), and a peak density of $\sim 3\times 10^6$\,cm$^{-3}$ at the core center.
It should be mentioned that we do find our RADEX column density estimates remain robust across our range of $n_{\mathrm{H}_2}$ input values (see Appendix\,\ref{appendix_radex_uncertain}).

The peak line temperatures ($T_\mathrm{mb}$ in units of K) and linewidths (FWHM in units of km/s) that are additional inputs for our RADEX calculations come directly from the Gaussian fits (Table\,\ref{tab:lines}). As shown in Figure\,\ref{fig:ch3oh}, there are multiple velocity components that we observe with the ARO 12\,m for the three separate CH$_3$OH 96.7\,GHz lines that are within $\sim$5\,MHz in frequency of each other, which complicates the line fitting. We find that a single component fit (top spectrum in Figure\,\ref{fig:ch3oh}) does not fully reproduce the observed spectrum (and neither does a two-component fit), instead a three-component velocity fit (bottom spectrum in Figure\,\ref{fig:ch3oh}) best reproduces the observed spectrum. The velocity components are comprised of a single centrally peaked component (i.e., at the v$_{lsr}$ of 3.8\,km/s) as well as two broader components that are blue-shifted and red-shifted from this main component. In the Yebes 40\,m data a single blue-shifted emission peak is also observed at $\sim$1.5\,km/s for the $1_{0,1}-0_{0,0}$\,A transition at 48.372\,GHz (see Figure\,\ref{fig:yebes_total}). This blue-shifted emission feature for CH$_3$OH is seen in higher transition lines (e.g., $5_{1,5}-4_{1,4}$ at 241.767\,GHz and $5_{0,5}-4_{0,4}$ at 241.791\,GHz) as reported in \cite{2024arXiv241213760S}. We conclude, as these authors do, that the emission we probe toward the IRAS 16293E peak therefore likely also contains the line-of-sight emission from the nearby `shocked gas' emission peaks, E1 and HE2 (Figure\,\ref{fig:map}; \citealt{2023A&A...673A.143K}). Because we can not disentangle this emission within our different beam sizes, and are limited in velocity resolution for our Yebes 40\,m data, we assume a single component fit around the centrally peaked component for all transitions observed and thus stress that here we provide a global average for the cold COM emission toward IRAS\,16293E. 

Beyond the physical conditions and line parameters that need to be inputted in the RADEX code, collisional rates are also necessary. In the case of CH$_3$OH, there are separate A and E states with measured collisional rates \citep{2010MNRAS.406...95R} and they are treated as two distinct molecules with H$_2$ as the collisional partner. Collisional rates for just the A-state transition of HCOOCH$_3$ are also available and were scaled from He to para-H$_2$ as the collisional partner \citep{2014ApJ...783...72F, 2020Atoms...8...15V}. {This data for CH$_3$OH and HCOOCH$_3$ comes from the LAMDA online database\footnote{\url{https://home.strw.leidenuniv.nl/~moldata/}} \citep{2005A&A...432..369S}, whereas for $^{13}$CH$_3$OH the EMAA Database\footnote{\url{https://emaa.osug.fr/species-list}} is used and provides the collisional data needed for RADEX where, again, para-H$_2$ is the collisional partner \citep{2024MNRAS.527.2209D}. }

We observed a total of three transitions for CH$_3$OH E and two transitions for CH$_3$OH A; therefore, using the RADEX code, {for a given $T_\mathrm{k}$ value, we run }a grid of $N$ values for all these transitions and then minimize over the observed and RADEX-{calculated peak line temperatures} to find best-fit $N$ values (see also \citealt{2020ApJ...891...73S, 2024MNRAS.533.4104S} {and Appendix\,\ref{appendix_radex_min}}). Next, we iterate for different $T_\mathrm{k}$ and $\theta_\mathrm{src}$ values to find a consistent best-fit $N$ for both A and E states\footnote{For CH$_3$OH A, it was more useful to iterate over $\theta_\mathrm{src}$ due to both transitions having different $\theta_\mathrm{beam}$ values. The CH$_3$OH E transitions all have roughly the same $\theta_\mathrm{beam}$ and therefore we used these `E' lines to constrain $T_\mathrm{k}$.}. We find our best-fit $N$ and $T_\mathrm{ex}$ values at $T_\mathrm{k}$ of 7.0\,K and a $\theta_\mathrm{src}$ of 115\,\arcsec (see Table\,\ref{table:colden}), revealing the emission is larger than our beam but $f$ does not equal one. For example, for an ARO 12\,m beam of $62\,\arcsec$ $f = 0.76$ and for an Yebes 40\,m beam of $37\,\arcsec$ $f = 0.90$, when $\theta_\mathrm{src}$ = 115\,\arcsec. These filling factors have non-negligible effects on the column density and hence why we are able to constrain $\theta_\mathrm{src}$ in the first place (see Appendix\,\ref{appendix_radex_min} for more discussion).

The total (A+E) column density found for CH$_3$OH is $N = (1.21 \pm 0.20) \times 10^{14}$\,cm$^{-2}$ at an average excitation temperature, $T_\mathrm{ex}$, of $6.6\pm0.4$\,K for a $\theta_\mathrm{src}$ of 115\,\arcsec (Table\,\ref{table:colden}). Our estimate for $N$ is higher by roughly an order of magnitude compared to LTE estimates in \cite{2023A&A...673A.143K} that find a $N$(CH$_{3}$OH) $= 8.0 \pm 1.6 \times 10^{12}\,\mathrm{cm}^{-2}$ toward IRAS\,16293E from higher energy transitions (in the $277-375$~GHz range with E$_u \sim 17-50$\,K), assuming a fixed source size of $20\arcsec$, an excitation temperature of $T_\mathrm{ex} = 12$\,K, and a 20\% error. We do note that if $T_\mathrm{ex}$ was lowered to $\sim7$\,K in an LTE analysis then \cite{2023A&A...673A.143K} would have obtained only a slightly higher estimate for the column density, $N \sim 1 \times 10^{13}$\,cm$^{-2}$, within the 20\% error. Of course the extended emission our observations probe is larger in size, and at $\theta_\mathrm{src}$ of 115\,$\arcsec$ this encompasses both E1 and HE2 emission peaks (see Figure\,\ref{fig:map}). 
The discrepancy in methanol column density could therefore be either a consequence of the contribution from E1 and HE2, or because our lower E$_u$ transitions (E$_u \sim 2-20$\,K) that we observe at lower frequencies, which are optically thin ($\tau = 0.03-0.55$ from RADEX), probe the cooler, less dense, and more extended methanol emission across the IRAS\,16293E prestellar core and surrounding cloud. The APEX observations from \cite{2023A&A...673A.143K} at higher frequencies selectively trace either higher density regions and/or hot regions, therefore it is likely our ARO 12\,m and Yebes 40\,m observations probe different excitation and physical conditions in the cloud. This is illustrated, for example, by the different CO maps in \cite{2023A&A...673A.143K}, which show the higher energy CO J=6-5 line ($E_\mathrm{u} = 116.2$\,K) tracing compact emission while the lower energy CO J=3-2 line ($E_\mathrm{u} = 33.2$\,K) traces more extended scales (even larger than 115\,arcsec; see their Figures 2 and J.51). Future work to map CH$_3$OH at intermediate frequencies (e.g., in the 2mm band) would help disentangle this inconsistency with ARO 12 m and APEX column density estimates.

The uncertainty on our estimate of the CH$_3$OH column density may be exacerbated due to the fact that only a single velocity component is fit to the complex line profiles (Figure\,\ref{fig:ch3oh}). We thus also calculate an $N$ for the more optically thin transitions of $^{13}$CH$_3$OH with less complex line profiles (Figure\,\ref{fig:12m_total}). We find a total (A+E) column density for $^{13}$CH$_3$OH of $N = (1.65 \pm 0.21) \times 10^{12}$\,cm$^{-2}$ at an excitation temperature of  $T_\mathrm{ex} \sim 6.6$\,K, and $\tau \sim 0.01$. Note that since we only have a single transition for each of the $^{13}$CH$_3$OH A and E states, respectively, we assume the same $T_\mathrm{k}$ of 7.0\,K and $\theta_\mathrm{src}$ of $115\arcsec$ that we find for CH$_3$OH. We can now calculate a $^{12}$C/$^{13}$C ratio and find a value of 73.4$\pm$9.51, which agrees within errors with the standard ISM $^{12}$C/$^{13}$C ratio of 68 \citep{2005ApJ...634.1126M}. We have successfully isolated the narrow component of the methanol emission, but are still unable to measure the $^{12}$C/$^{13}$C ratio for blue- and red-shifted components of the CH$_3$OH emission that do not appear in the  $^{13}$CH$_3$OH spectra (Figures\,\ref{fig:ch3oh} and \ref{fig:12m_total}).

Next, for HCOOCH$_3$ A, we minimize over the two observed transitions available to find that a $T_\mathrm{k}$ of 27\,K is a best-fit to the observations (see Appendix\,\ref{appendix_radex_min}). A total (A+E) $N$ value of $(4.9\pm0.1) \times 10^{12}$\,cm$^{-3}$ at an average  T$_\mathrm{ex}$ of 26.0$\pm$4.59\,K is calculated (Table\,\ref{table:colden}). The larger uncertainty in T$_\mathrm{ex}$ (compared to the methanol species) stems from the fact that these species are in non-LTE, i.e., the RADEX-derived $T_\mathrm{ex}$ for each transition of HCOOCH$_3$ A differs by $\sim5-10$\,K at a single given $T_\mathrm{k}$. Both of our best-fit $T_\mathrm{k}$ and T$_\mathrm{ex}$ values should be treated with caution, due to the uncertainty of the 100\,GHz $8_{1,7} - 7_{1,6}$ A line detected at the $3\sigma$ limit, which has been smoothed by three channels and is significantly broader than the corresponding 36\,GHz $3_{0,3} - 2_{0,2}$ A line (Table\,\ref{tab:lines}). If the $8_{1,7} - 7_{1,6}$ A transition is treated as a non-detection, the best-fit column density constraint remains robust but the best-fit kinetic temperature range can now also vary by more than a factor of two, from $\sim14-30$\,K (see Figure\,\ref{fig:RADEX_grid_HCOOCH3} in Appendix\,\ref{appendix_radex_min}). Our calculation also assumes an A:E of 1:1, which we expect to be reasonable within our error because for CH$_3$OH, the $N$ values for the separate A and E states were within a factor of 1.04. As we only have two transitions for HCOOCH$_3$ A and we already minimized over $T_\mathrm{k}$, we cannot constrain $\theta_\mathrm{src}$ and assume $f = 1$. 

\subsection{{LTE method: CH$_2$DOH, CHD$_2$OH, CH$_3$CHO, CH$_3$OCH$_3$ and CH$_2$CHCN}} \label{sec:subLTE}

For the deuterated methanol species (CH$_2$DOH and CHD$_2$OH), as well as the other COMs targeted (CH$_3$CHO, CH$_3$OCH$_3$ and CH$_2$CHCN), we must use the Rotation Diagram (RD) method or the Local Thermodynamical Equilibrium (LTE) method to calculate $N$ as collisional coefficients needed for the non-LTE analysis are not available. Under this approximation, the upper state column density in the optically thin limit can be calculated by: \begin{equation}\label{eq:Nu}
    \mathrm{N}_u = \frac{I}{h A_{ul} f} \frac{u_\nu(\mathrm{T}_\mathrm{ex})}{[J_\nu(\mathrm{T}_\mathrm{ex}) - J_\nu(\mathrm{T}_\mathrm{cmb})]},
\end{equation} where $h$ is the Planck constant, $f$ is our (frequency-dependent) filling factor, $I$ is the integrated intensity of the line, $A_{ul}$ is the spontaneous emission coefficient (or `Einstein A'), $T_\mathrm{cmb}$ is the background temperature of 2.73\,K, and $u_\nu$ (Planck energy density) and $J_\nu$ (Planck function in temperature units) defined as: \begin{equation}
    u_\nu \equiv \frac{8 \pi h \nu^3 }{c^3} \frac{1}{\exp{(h\nu/kT) - 1}}, 
\end{equation} \begin{equation}
    J_\nu \equiv \frac{h \nu }{k} \frac{1}{\exp{(h\nu/kT) - 1}}.
\end{equation} In these equations, $c$ is the speed of light, $k$ is the Boltzmann constant, and $\nu$ is our line frequency. For a total column density, $\mathrm{N_{tot}}$, \begin{equation} \label{eq:N}
    \frac{\mathrm{N}_u}{g_u} = \frac{\mathrm{N_{tot}}}{\mathrm{Q(T_{ex})}} \exp{(-E_u / k \mathrm{T_{ex}})},
\end{equation} where $g_u$ is the upper state degeneracy, $E_u$ is the upper state energy, and $\mathrm{Q(T_{ex})}$ is the partition function dependent on the excitation temperature of the molecule. In the standard RD method, the log-normal of the left side of equation\,\ref{eq:N}, ln(${N}_u/g_u$), is plotted versus $E_u$ so that the excitation temperature, ${T_\mathrm{ex}}$, is the inverse of the slope of the linear fit and the y-intercept is used to find ${N_\mathrm{tot}}$ \citep{1999ApJ...517..209G}. In our estimates both $J_\nu$(T$_\mathrm{ex}$) and $J_\nu$(T$_\mathrm{cmb}$) in equation\,\ref{eq:Nu} are considered given that in the lower $T_\mathrm{ex}$ limit, the contribution of $J_\nu$(T$_\mathrm{cmb}$) is non-negligible. We then perform an iterative fitting where all inputs are fixed except the input $T_\mathrm{ex}$, which is optimized when it equals the inverse slope of the rotation diagram. We use this approximation to find $N$ and $T_\mathrm{ex}$ for CH$_3$CHO A and CH$_2$DOH e$_0$ (Figure\,\ref{fig:rotatation}). We note that $\mathrm{Q(T_{ex})}$ is calculated separately for the individual torsional level sub-states, i.e., `A' for CH$_3$CHO and `EE' for CH$_3$OCH$_3$ (see Appendix\,A in \citealt{2023PhDT........72S} for details).

\begin{table}
\caption{Molecular column densities in IRAS 16293E}  
\setlength{\tabcolsep}{4pt}
\label{table:colden}                           
\begin{tabular}{l l c c c c c c c c }     
\hline\hline            
Method & Molecule  & $T_\mathrm{ex}$   &  $N$  \\ [1pt] 
        &  & (K)  &   (cm$^{-2}$) &   \\ [3pt]
\hline 
RADEX &       CH$_3$OH   E    &   6.54$\pm$0.32  & (6.17$\pm$1.95) $\times$ 10$^{13}$  \\ [3pt]
&       CH$_3$OH   A    &   6.60$\pm$0.47  & (5.94$\pm$0.20) $\times$ 10$^{13}$  \\ [3pt]
& $^{13}$CH$_3$OH E     &   6.75$\pm$1.34  & (8.00$\pm$1.50) $\times$ 10$^{11}$ \\ [3pt]
& $^{13}$CH$_3$OH A      &   6.64$\pm$1.28  & (8.50$\pm$0.15) $\times$ 10$^{11}$ \\ [3pt]
& HCOOCH$_3$  A    & 26.0$\pm$4.59  & (2.45$\pm$0.05) $\times$ 10$^{12}$  \\ [3pt]
\hline 
RD & CH$_3$CHO  A  & { 6.81$^{+0.33}_{-0.60}$} & {1.84$^{+0.49}_{-0.48}$ $\times$ 10$^{12}$} \\ [3pt]
&  CH$_2$DOH  e$_0$     & {4.55$^{+0.20}_{-0.36}$} & {7.52$^{+2.18}_{-2.17}$ $\times$ 10$^{12}$} \\ [3pt]
\hline
LTE & CHD$_2$OH e$_0$  & {4.55} &  {(2.64$\pm$0.83)\,$\times$ 10$^{12}$}\\ [3pt]
& {CH$_3$OCH$_3$ EE} & {10} & {(1.30$\pm{0.39})$ $\times$ 10$^{12}$} \\
& CH$_2$CHCN & 10 &  $<$0.05\,$\times$ 10$^{12}$\\ 
\hline                                    
\end{tabular}
\tablefoot{The best-fit kinetic temperature, $T_\mathrm{k}$, used in the RADEX calculations for CH$_3$OH is 7.0\,K (also used for $^{13}$CH$_3$OH) and for HCOOCH$_3$ A it is 27\,K.      For all methanol species and CH$_3$CHO a source size, $\theta_\mathrm{src}$, of $115\arcsec$ is used. In the LTE method, we assume a fixed $T_\mathrm{ex}$ with no associated error. Here `RD' denotes for which molecules the Rotation Diagram method was used.} 
\end{table}

\begin{figure*}
   \centering
    \includegraphics[width=\textwidth]{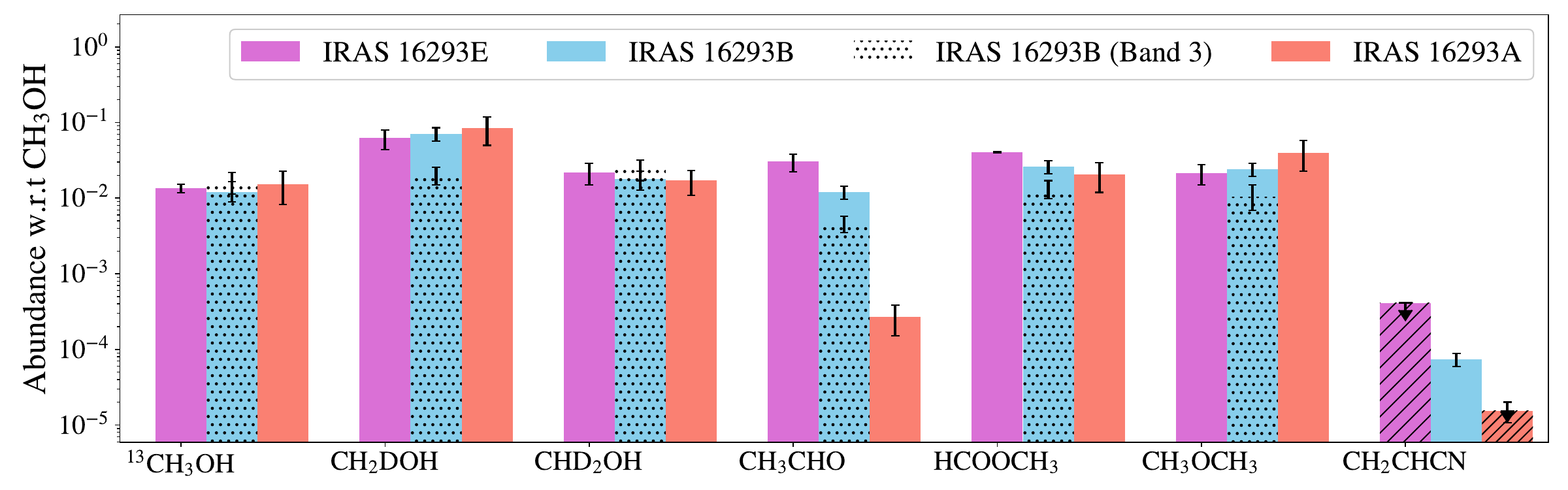}
    \caption{Comparison of D-COM and COM abundances normalized to methanol for IRAS 16293E and the nearby protostars A and B, from ALMA Band 7 PILS data (\cite{2018A&A...620A.170J, 2018A&A...616A..90C, 2020A&A...635A..48M, 2022A&A...659A..69D}), as well as from \cite{2024A&A...686A..59N} that observed IRAS\,16293B in ALMA Band 3 (overlaid dotted pattern). Upper limits for CH$_2$CHCN are bins with downward arrows that have an overlaid hatching pattern (note: not shown is the IRAS\,16293B Band 3 upper limit for CH$_2$CHCN, which is $< 4.9 \times 10^{-4}$). 
    Note that the methanol column density for IRAS~16293B (Band 7) stems from the analysis of the optically thin $^{18}$O-isotopologue of methanol and subsequent assumption of the canonical $^{16}$O/$^{18}$O ISM ratio \citep{2018A&A...620A.170J}, while that for IRAS~16293A is based on the analysis of carefully selected optically thin lines of the main isotopologue itself \citep{2020A&A...635A..48M}.} 
    \label{fig:meth_abund}
\end{figure*}

In the case of CH$_3$CHO A, there are five transitions that are used in the calculation. While the ARO 12\,m transitions show direct evidence that multiple velocity components are present (Figure\,\ref{fig:12m_total}), we cannot resolve this in our Yebes 40\,m observations due to the lower velocity resolution (Figure\,\ref{fig:yebes_total}) and therefore consider only single-component Gaussian fits. The best-fit to the data finds an $N = (1.84^{+0.49}_{-0.48}) \times 10^{12}$\,cm$^{-2}$ at a $T_\mathrm{ex} \sim 6.8$~K, which is in agreement with the $T_\mathrm{ex}$ calculated for CH$_3$OH (Table\,\ref{table:colden}). For CH$_3$CHO A, we also assume that it is spatially extended similarly to CH$_3$OH (as done in \citealt{2021MNRAS.504.5754S, 2024MNRAS.533.4104S}) and thus correct for $\theta_\mathrm{src} = 115\arcsec$. If, instead, $f =1$ is assumed, then our $N$ is only slightly decreased, by roughly $10\%$ and within our calculated errors. A total (A+E) column density is therefore $N = (3.68^{+0.97}_{-0.95}) \times 10^{12}$\,cm$^{-2}$ and assumes (as done above for HCOOCH$_3$) an A:E ratio of 1:1. 

For CH$_2$DOH, we were able to observe three distinct more reliable a-type transitions (where $\Delta$K$_a = 0, \pm2, ...$), two with the ARO 12\,m and one with the Yebes 40\,m (see Table\,\ref{tab:lines}), so a rotation diagram was constructed and both $N$ and $T_\mathrm{ex}$ constrained. As was the case for CH$_3$CHO, we assume $\theta_\mathrm{src} = 115\arcsec$ to find
{$N = (7.52^{+2.18}_{-2.17})$ $\times 10^{12}$\,cm$^{-2}$ at a $T_\mathrm{ex}$ of $4.55^{+0.20}_{-0.36}$\,K. If} $f = 1$ is assumed, $N$ decreases by $< 10\%$.

For CHD$_2$OH, only two transitions were observed and their $E_\mathrm{u}$ values only differ by $\sim$3\,K (Table\,\ref{tab:lines}); so we assume the same T$_\mathrm{ex}$ found for CH$_2$DOH of 4.55\,K. This is the `LTE' method, which we point out is just an assumption that must be made, and likely T$_\mathrm{k}$ is larger than T$_\mathrm{ex}$ here. Assuming again that $\theta_\mathrm{src} = 115\arcsec$, $N = (2.64\pm0.83) \times 10^{12}$\,cm$^{-2}$ is obtained for CHD$_2$OH. The LTE method is also used to estimate an upper limit for CH$_2$CHCN at a fixed $T_\mathrm{ex}$ of 10\,K (see Table\,\ref{table:colden}).

As can be seen in Figure\,\ref{fig:yebes_total}, the EE state of CH$_3$OCH$_3$ is the brightest, and detected with the Yebes 40\,m at 32.9\,GHz ($E_\mathrm{u} = 7.0$~K) and 39.0\,GHz ($E_\mathrm{u} = 15.4$~K). Additionally, there is a $3\sigma$ upper limit provided from the ARO 12\,m at 99\,GHz ($E_\mathrm{u} = 10.2$~K), whose $rms$ level was $\sim3\times$ worse preventing a firm detection (see Table\,\ref{tab:lines} and Figure\,\ref{fig:12m_total}). While these three constraints from three separate transitions can be used to create a RD (see Appendix\,\ref{appendix_LTE}), several limitations in the data make this impractical. Mainly, the 39.0\,GHz $5_{1,4}-5_{0,5}$ transition may be untrustworthy due to 1) the other AA and AE+EA states not being detected, 2) the slight 0.4\,km/s offset in the v$_{lsr}$ and 3) the negative feature at 2.7\,km/s as deep as the detected emission line (Figure\,\ref{fig:yebes_total}). Therefore, instead we assume a standard $T_\mathrm{ex}$ value of 10\,K and calculate $N = 1.30\pm0.39 \times 10^{12}$\,cm$^{-2}$. We note that unlike for CH$_3$CHO and the deuterated methanol species, we are made to assume here $f = 1$ (as done for HCOOCH$_3$) because we do not have any estimate of the emitting area for this molecule.

If the 39.0\,GHz line of CH$_3$OCH$_3$ is a `true' detection, then this would indicate an elevated excitation temperature  ($\sim30$\,K; see Appendix\,\ref{appendix_LTE}). It is plausible that the $T_\mathrm{ex}$ for both HCOOCH$_3$ and CH$_3$OCH$_3$ is enhanced compared to the methanol species and CH$_3$CHO (Table\,\ref{table:colden}), indicating that both species {may be} tracing gas heated by the shocks from the nearby protostars (see Section\,\ref{sec:discuss} for more discussion). {But, due to the limited number of transitions and low signal-to-noise level of the data, we are unable to confirm this temperature enhancement. }

\begin{table}
\caption{Methanol deuterium ratios in IRAS 16293-2422}  
\setlength{\tabcolsep}{9pt}
\label{table:deut}    
\begin{tabular}{r c c c }      
\hline\hline             
 &  \multicolumn{3}{c}{D/H (\%) }  \\ [1pt] 
   &  16293A &  16293B &  16293E  \\ [1pt] 
\hline 
  CH$_2$DOH/CH$_3$OH  & 2.8$\pm$1.2 & 2.4$\pm$0.5 &  {2.1$\pm$0.6}  \\ [2pt]
 CHD$_2$OH/CH$_3$OH & 7.5$\pm$1.1 &  7.7$\pm$1.2  & {8.5$\pm$2.7}\\ [2pt]
\hline                \hline              
     &  16293A &  16293B &  16293E  \\ [2pt] 
 \hline 
  CHD$_2$OH/CH$_2$DOH & 20$\pm$6.8 & 25$\pm$5.0 &  {35$\pm$15}\\ 
  \hline 
\end{tabular}
\tablefoot{Statistical corrections applied to the D/H ratio {(top two rows)} where CH$_2$DOH/CH$_3$OH =3(D/H) and CHD$_2$OH/CH$_3$OH=3(D/H)$^2$ (appendix C of \citealt{2022A&A...659A..69D} and appendix B of \citealt{2019A&A...623A..69M}). Values reported here for IRAS 16293B from the Band 7 PILS observations \citep{2018A&A...620A.170J}.}
\end{table}

\section{Discussion} \label{sec:discuss}

The prestellar core IRAS\,16293E gives us a direct view into the nascent environment where the chemically rich and well-studied protostars IRAS\,16293A and B also formed. Below we compare our derived prestellar abundance ratios while putting into context our observations with other star-forming regions and later stage objects (e.g., protostars and comets). 

\subsection{Abundance comparisons}

We first normalized our IRAS\,16293E calculated COM and D-COM column densities to the CH$_3$OH column density and compared them to the nearby protostars IRAS\,16293A and B (Figure\,\ref{fig:meth_abund}). The protostellar data comes both from the higher frequency (275-373\,GHz) ALMA Band 7 PILS data \citep{2018A&A...620A.170J, 2018A&A...616A..90C, 2020A&A...635A..48M, 2022A&A...659A..69D}, as well as from lower frequency (84-116\,GHz) ALMA Band 3 data solely for IRAS\,16293~B \citep{2024A&A...686A..59N}. For these high temperature and high density protostellar regions one needs to worry more about optically thick methanol, and so it should be noted that the methanol column density for IRAS~16293B in the Band 7 data \citep{2018A&A...620A.170J} stems from the analysis of the optically thin $^{18}$O-isotopologue of methanol and subsequent assumption of the canonical $^{16}$O/$^{18}$O ISM ratio of 560 \citep{1994ARA&A..32..191W}, while that for IRAS~16293A is based on the analysis of carefully selected optically thin lines of the main isotopologue itself \citep{2020A&A...635A..48M}. In the recent Band 3 observations \citep{2024A&A...686A..59N}, $^{13}$CH$_3$OH is scaled to the main isotopologue by the $^{12}$C/$^{13}$C ISM value of 68 \citep{2005ApJ...634.1126M}.

By normalizing to CH$_3$OH, which primarily forms on the grains \citep{2002ApJ...571L.173W, 2009A&A...505..629F}, we can use the relative abundances of the other species to test if the initial inventory of material survived the star formation process. 
As seen in other studies that compare larger heterogeneous samples (e.g., \citealt{2020A&A...639A..87V, 2021A&A...650A.150N, 2021MNRAS.504.5754S, 2024MNRAS.533.4104S}), we find that the protostellar COM and D-COM relative ratios with respect to CH$_3$OH in IRAS\,16293~A and B are similar those in the prestellar core IRAS\,16293E.  
Unlike other studies, however, we uniquely compare these different stages of star formation that have evolved from the same common molecular cloud and thus show directly the inheritance of COMs from the prestellar stage. 

Particularly interesting to point out in our comparison are the deuteration ratios, which in the case of CHD$_2$OH, are statistically identical for IRAS\,16293A, B, and E. We investigate this further by calculating the D/H ratios accounting for the statistical corrections that consider the number of indistinguishable combinations with $i$ deuterium atoms at $n$ potential sites in a specific functional group (Table\,\ref{table:deut}).  
The D/H ratios for both CH$_2$DOH and CHD$_2$OH agree within error with a slight increase and decrease, respectively, from IRAS\,16293E to B to A. Correspondingly, the CHD$_2$OH/CH$_2$DOH ratios agree within error but show a downward trend as cores evolve, i.e., for IRAS 16293E, B, and A the CHD$_2$OH/CH$_2$DOH ratios are 35$\pm15\%$, 25$\pm5.0\%$ and 20$\pm6.8\%$, respectively. 
Another grain species, water, provides additional insight into the chemical evolution of these cores via deuteration, as \cite{2004ApJ...608..341S} found similar HDO abundances in IRAS 16293E and A suggesting that the warm gas in IRAS 16293A may have undergone the same evolution and pre-collapse phase that IRAS 16293E is currently in. We therefore show directly from our D/H ratios that the enhanced deuterated methanol in protostars IRAS 16293A and B was set during the prestellar phase. 

We also briefly mention the order-of-magnitude difference in the CH$_3$CHO abundance from IRAS\,16293A to IRAS\,16293B, which has been discussed in \cite{2020A&A...635A..48M} as due to differences in the spatial distributions of the molecule at different excitation conditions enhanced by the differences in geometry of the two protostellar sources.
The CH$_3$CHO abundance we measure for the prestellar source IRAS\,16293E is much more consistent with IRAS\,16293B, yet still enhanced when compared to both protostellar sources (16293B and 16293A; see Figure\,\ref{fig:meth_abund}). A higher abundance toward IRAS\,16293E supports the formation scenario where CH$_3$CHO is forming first in the gas phase in the early cold core stage (e.g., \citealt{2013ApJ...765...60G, 2020MNRAS.499.5547V, 2022ApJS..259....1G}). 

\subsection{Influence from nearby outflows}

The outflow interaction from the IRAS\,16293-2422 protostellar complex has been traced, for example, by rotational transitions of molecular hydrogen picked up by 4.5~$\mu$m Spitzer observations (see Figure 8 in \citealt{2008ApJ...683..822J}). This emission lies spatially above IRAS\,16293E and roughly matches the higher frequency CH$_3$OH emission as mapped by \cite{2023A&A...673A.143K} and plotted as contours in Figure\,\ref{fig:map}. As shown in this map, the observed beam sizes from our single-pointing observations clearly overlap with this outflow-related emission and this results in the observed multiple velocity components for CH$_3$OH and CH$_3$CHO, as discussed above (Section\,\ref{sec:results}), as well as the subsequent warming of the surrounding prestellar envelope gas that could have allowed for the enhancement of observed gas phase COM abundances toward IRAS 16293E, which we discuss here. 

The abundances relative to methanol of the largest COMs we targeted in IRAS\,16293E, HCOOCH$_3$ and CH$_3$OCH$_3$, are comparable to the IRAS\,16293A and B protostars (Figure\,\ref{fig:meth_abund}). We find evidence that HCOOCH$_3$ and CH$_3$OCH$_3$ may potentially even  trace warmer gas in the envelope surrounding the central `E' core that has been influenced by outflow emission. In the case of HCOOCH$_3$ A, both $T_\mathrm{ex}$ and $T_\mathrm{k}$ are found to be $\sim27$\,K in our RADEX calculations (assuming the secure detection of both A state transitions listed in Table\,\ref{tab:lines}), though the range can expand to $T_\mathrm{k}$$\sim 14 -30$\,K if only the 36\,GHz line is considered trustworthy (Appendix\,\ref{appendix_radex_min}). As for CH$_3$OCH$_3$ EE, a $T_\mathrm{ex}$ of $\sim$30\,K is found through the LTE RD method if the 39\,GHz line is considered trustworthy (Figure\,\ref{fig:dme_rot} in Appendix\,\ref{appendix_LTE}), causing a $\times3$ enhancement in abundance compared to the fixed $T_\mathrm{ex} = 10$\,K estimate in Table\,\ref{table:colden}.  Methanol species and CH$_3$CHO, in contrast, show $T_\mathrm{ex} \sim 5-7$\,K and for CH$_3$OH in particular a $T_\mathrm{k}$ of $7$\,K is found. 

\cite{2024MNRAS.533.4104S} did a similar analysis as we present in Section\,\ref{sec:results} {(and Appendix\,\ref{appendix_radex_min})} to find the best-fit $T_\mathrm{k}$ for HCOOCH$_3$ A from {minimizing a RADEX grid} for a sample of starless and prestellar cores in the Perseus Molecular Cloud. That work showed that for prestellar and starless cores in regions with active protostellar outflows the $T_\mathrm{k}$ for  HCOOCH$_3$ is also enhanced, around 20\,K. In contrast, the best-fit $T_\mathrm{k}$ values for CH$_3$OH in this sample of cores is $< 10$\,K (see \citealt{2024MNRAS.533.4104S} Appendix D). While the small number of detected transitions at low signal-to-noise levels limit our confidence that $T_\mathrm{ex}$ or $T_\mathrm{k}$ are truly enhanced for HCOOCH$_3$ and/or CH$_3$OCH$_3$ toward prestellar sources in shocked regions, the presence or history of shocks in these regions may have still helped to facilitate the production of gas-phase COMs.

\begin{figure*}
   \centering
    \includegraphics[width=160mm]{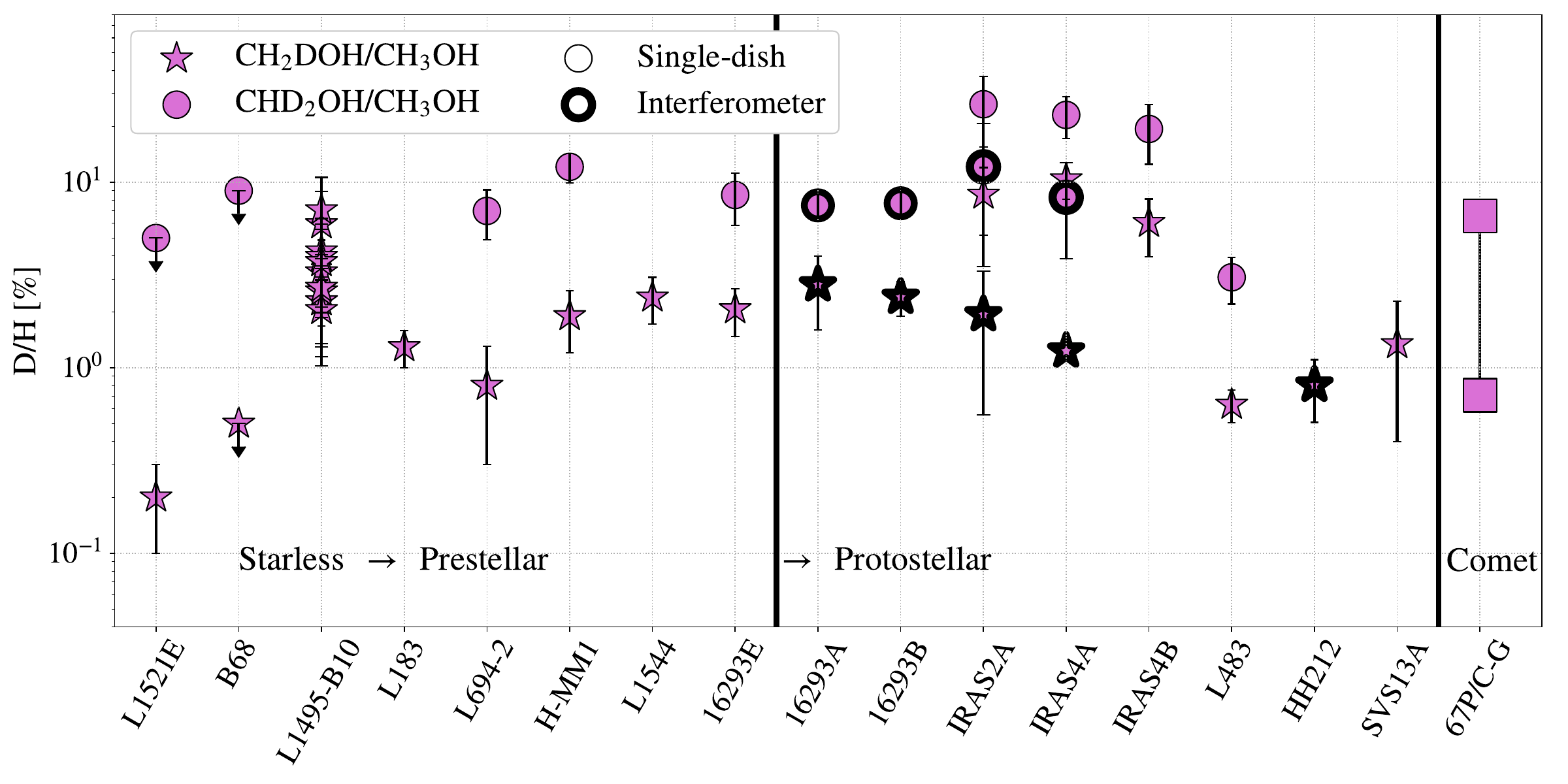}
    \caption{The statistically corrected D/H ratios for all starless and prestellar sources where data is available \citep{2019A&A...622A.141C, 2020A&A...633A.118L, 2021MNRAS.501..347A, 2023A&A...669L...6L}, including IRAS 16293E presented in this work, as well as for a mix of single-dish and interferometric (thicker outlined markers) measurements toward protostars \citep{2006A&A...453..949P, 2017MNRAS.467.3011B, 2017A&A...606L...7B, 2019A&A...625A.147A, 2019A&A...632A..19T} and the 67P/Churyumov–Gerasimenko (67P/C-G) comet, which is shown instead as a range of values (Table 1 in {\citealt{2021MNRAS.500.4901D}}) because these measurements are from the ROSINA mass spectrometer and thus the isotopologues are not uniquely identified and more uncertain than the spectroscopic data points. Upper limits are shown with downward arrows. See also \cite{2022A&A...659A..69D} and \cite{2023A&A...669L...6L} for similar comparison plots. 
    } 
    \label{fig:dh_ratio}  
\end{figure*}

Even before COM detections in prestellar sources, the first detection of HCOOCH$_3$ (among other species) toward the active outflow L1157-B1 revealed that shock-induced sputtering can work to help release these larger molecules from the grains \citep{2008ApJ...681L..21A}. Later, \cite{2010ApJ...716..825O} reported the detections of the COMs CH$_3$CHO, HCOOCH$_3$, and (tentatively) CH$_3$OCH$_3$, toward the B1-b system that includes a protostar, outflow, and dense core. To date, the influence of outflows and shocks and subsequent formation of complex chemistry directly in and around low-mass prestellar sources still remains much of a mystery. 

In the high-mass case, the dark cloud G+0.693-0.027 is thought to have been shocked (e.g., \citealt{2024A&A...690A.121C}) and provides the evidence for why many highly complex species, including the recent detections of three C$_4$H$_3$N isomers (cyanoallene, CH$_2$CCHCN; propargyl cyanide, HCCCH$_2$CN; and cyanopropyne CH$_3$CCCN; \citealt{2022FrASS...9.6870R}) as well as the glycine isomer, glycolamide (NH$_2$C(O)CH$_2$OH; \citealt{2023ApJ...953L..20R}), have all been observed in the gas phase.

The detection of large molecules in regions of active outflows and/or previously shocked quiescent gas suggests that the production of COMs can proceed efficiently in environments where molecules get released into the gas phase from the ice mantle of the grains (see reviews \citealt{2020ARA&A..58..727J, 2023ASPC..534..379C}). Slow shocks from the IRAS 16293-2422 protostellar complex  (e.g., $v_{s}< 20$\,km/s; \citealt{2002ApJ...567..980G}) could therefore have released molecules, either from enhanced dust temperatures and/or ice sputtering, inducing more rapid COM formation. Our single-dish observations could be picking up enhanced HCOOCH$_3$ and CH$_3$OCH$_3$ abundances and excitation temperatures in this shocked warm cloud surrounding the core, while CH$_3$OH and CH$_3$CHO may more effectively trace the core itself as well as extended lower density and lower temperature gas with $T_\mathrm{k} \sim 7$\,K. This derived kinetic temperature for methanol ($7$\,K) is consistent with tracing the central prestellar core modeled to have $T_\mathrm{k} \sim 8$\,K via millimeter interferometric observations \citep{2001AAS...199.6001S, 2004ApJ...608..341S}.

As discussed in section\,\ref{sec:subRADEX}, it was not possible to derive an estimate on the source size for HCOOCH$_3$ and CH$_3$OCH$_3$, and therefore $f = 1$ is assumed. We do not know the emitting area of these higher complexity COMs, and therefore future spatial mapping of these low frequency transitions is needed to confirm if they trace emission in the extended cloud of IRAS\,16293E. And, while we do fit a source size for CH$_3$OH of $\theta_\mathrm{src} = 115\arcsec$, and assume the same value for its isotopologues and the chemically related CH$_3$CHO, we caution that this may not be a true representation of the emitting region for methanol, whose kinetic temperature ($\sim 7$\,K) would indicate it's tracing the central prestellar core. The complicated nature of the region suggests possible contamination from positions E1 or HE2 (Figure\,\ref{fig:map}).

\subsection{Complex chemistry in context with other cores}

The statistically corrected D/H ratios for the deuterated methanol species (CH$_2$DOH/CH$_3$OH and CHD$_2$OH/CH$_3$OH) are plotted in Figure\,\ref{fig:dh_ratio} for all starless and prestellar cores where data is available, as well as for a sample of protostars (both single-dish and interferometric observations) and the comet 67P/Churyumov–Gerasimenko (67P/C-G). In general, we see what one might expect, the D/H ratio enhancing from starless cores to prestellar cores, where lower temperatures and higher CO-depletion factors favor deuterium enrichment (e.g., \citealt{2002P&SS...50.1133C, 2004A&A...418.1035W, 2005ApJ...619..379C, 2005A&A...438..585R}).
The enhanced ratio is then maintained into the protostellar stages; and these values match up well with the D/H range for comet 67P/C-G {\citep{2021MNRAS.500.4901D}}. Additionally, these ratios show a higher D/H ratio of the doubly deuterated molecules than the singly deuterated ones due to more efficient deuteration once one D substitution has already taken place. This successive deuteration, i.e., higher levels of fractionation in multi-deuterated isotopologues, has been studied previously (e.g., see \citealt{2022A&A...659A..69D}). The similarities in D/H ratios across evolutionary stages for methanol deuteration have been seen before in \cite{2023A&A...669L...6L}, and confirm a chemical inheritance from the prestellar stage. 

In particular, we look at where IRAS\,16293E falls in relation to these literature values to find tight agreement. Both the CH$_2$DOH/CH$_3$OH and CHD$_2$OH/CH$_3$OH values in IRAS\,16293E are within a factors of $\sim1-3$ to the prestellar sources H-MM1, which is also located in Ophiuchus, and L694-2, which is more isolated \citep{2023A&A...669L...6L}. Additionally, the D/H ratio for the `prototypical' prestellar core L1544 \citep{2019A&A...622A.141C} and L183 prestellar core \citep{2020A&A...633A.118L} in the singly deuterated case also fall within a factor of 1.5 when compared to IRAS\,16293E. 

When comparing to the protostellar sources, we find in general that the interferometric protostellar observations are more consistent with the single-dish prestellar constraints, compared to the single-dish protostellar observations (Figure\,\ref{fig:dh_ratio} {and Table\,\ref{tab:compare_cores}}). The single-dish measurement could be underestimating the D/H ratio either due to beam dilution effects, evident from the $21-30^{''}$ beam observations of L483 \citep{2019A&A...625A.147A}, or due to a real evolutionary effect as gas phase deuterated methanol heats up and decreases in abundance during later star formation, as is the case for the only Class I source plotted, SVS13A (beam $\sim10^{''}$; \citealt{2017MNRAS.467.3011B}). 
In other cases, single-dish measurements may be overestimating the D/H ratio due to additional uncertainties, such as unreliable line data. For example, in \citealt{2016A&A...587A..91B} (see their footnote 7), it was reported that the CH$_2$DOH column densities from \cite{2002A&A...393L..49P} and \cite{2006A&A...453..949P} were overestimated by a factor of two, because the laboratory line intensities used were a factor of two lower than those now reported in updated catalogs. While we correct for this factor in our estimates for IRAS2A, IRAS4A, and IRAS4B in Figure\,\ref{fig:dh_ratio}, there may be additional factors of uncertainty from these early single-dish measurements and exact comparisons should be taken with caution. Future observations of deuterated methanol at higher spatial scales in both prestellar and protostellar cores should help to resolve these discrepancies. 

We also mention that, while not a low-mass source, new observations of CH$_3$OH and CH$_2$DOH in the Herbig Ae disk HD 100453 find a D/H ratio of $\sim 1-2\%$ \citep{2025ApJ...986L...9B}, consistent with the low-mass measurements complied in Figure\,\ref{fig:dh_ratio}. This first (tentative) detection of deuterated methanol in a Class II disk further adds to the growing evidence that organic material from the earliest stages are inherited to the later stages of star and planet formation.

The COMs targeted in this survey (CH$_3$OH, CH$_3$CHO, CH$_2$CHCN, HCOOCH$_3$, and CH$_3$OCH$_3$) are those regularly searched for in other starless and prestellar cores in the literature (e.g., \cite{2012A&A...541L..12B, 2014ApJ...795L...2V, 2016ApJ...830L...6J, 2020ApJ...891...73S, 2021MNRAS.504.5754S, 2021ApJ...917...44J, 2023MNRAS.519.1601M, 2024MNRAS.533.4104S}; see Table\,\ref{tab:compare_cores}). In this context, we suggest IRAS\,16293E is a `typically complex prestellar core' and compare its COM detections to the larger representative samples of cores in the Taurus L1495 filament  \citep{2020ApJ...891...73S, 2023PhDT........72S} and Perseus Molecular Cloud \citep{2024MNRAS.533.4104S}, with CH$_3$CHO detection rates $>50\%$ and CH$_2$CHCN detection rates of $<34\%$. The non-detection of CH$_2$CHCN in IRAS\,16293E, despite its substantial abundance in both the evolved prestellar core L1544 \citep{2016ApJ...830L...6J} and young starless core L1521E \cite{2021MNRAS.504.5754S}, may suggest that local environmental effects that have suppressed its gas phase abundance, but more investigation into its formation pathways in cold cores is needed (see discussion in \citealt{2024MNRAS.533.4104S}). It remains to be studied if IRAS\,16293E harbors any higher complexity species, which could be released into the gas phase due to the influence from the neighboring outflows and shocks.

\section{Conclusions} \label{sec:conclude}

IRAS 16293E is an intriguing prestellar core to study because it has some of the highest levels of deuteration among starless and prestellar cores and is believed to be forming in the same natal cloud environment as its chemically-rich protostellar neighbors, IRAS 16293A and B, formed in. In our joint ARO 12m and Yebes 40\,m survey, we detected for the first time $^{13}$CH$_3$OH, CH$_2$DOH, CHD$_2$OH, CH$_3$CHO, HCOOCH$_3$, and CH$_3$OCH$_3$ toward the emission peak of IRAS\,16293E that, when normalized to CH$_3$OH, show similar abundances when compared to IRAS 16293A and B. Our main findings are summarized below.

\begin{enumerate}

    \item By detecting two or more transitions of each COM or D-COM, we were able to use radiative transfer techniques to place tight constraints on abundances and excitation conditions. We found via the non-LTE RADEX method total (A+E) CH$_3$OH and $^{13}$CH$_3$OH column densities, $N$, of (1.21$\pm$0.20)$\times$10$^{14}$\,cm$^{-2}$ and (1.6$\pm$0.21)$\times$10$^{12}$\,cm$^{-2}$, respectively, and $T_\mathrm{ex}$ values ranging from 6.54 to 6.75\,K (for a $T_\mathrm{k}$ of 7.0\,K). These results corresponds to a $^{12}$C/$^{13}$C ratio of 73.4$\pm$9.51, which is in agreement with local ISM values.
    For CH$_3$CHO (and from the independent RD method), a similar $T_\mathrm{ex}$ of 6.81$^{+0.31}_{-0.60}$\,K is found, where the total column density is $N$\,=\,$(3.68^{+0.97}_{-0.95})\,\times\,10^{12}$\,cm$^{-2}$. While the $T_\mathrm{ex}$ values for higher complexity molecules HCOOCH$_3$ and CH$_3$OCH$_3$ are more uncertain, the total $N$ values are comparable to CH$_3$CHO (within {a factor of 2}) at (4.9$\pm$0.10)$\times$10$^{12}$\,cm$^{-2}$ and {(2.6$\pm$0.78)$\times$10$^{12}$\,cm$^{-2}$}, respectively.
    
    \item Influence from the nearby protostellar outflows onto IRAS\,16293E has warmed the outer envelope surrounding the central prestellar core and this emission is being probed within our single-dish beams (ranging from $37\arcsec$ to $73\arcsec$). The single, dominant emission component we analyze for CH$_3$OH and CH$_3$CHO traces the colder ($\sim$10\,K) gas in and around the core. Still, the fact that we observe multiple velocity components for CH$_3$OH and CH$_3$CHO in the higher velocity resolution ARO 12m observations (that are also larger in beam size and therefore encompass more of the outer envelope) provides additional evidence that different emission regions are probed. It is also possible that HCOOCH$_3$ and CH$_3$OCH$_3$ could be tracing the warmer envelope, where shock-induced ice sputtering as well as the additional production and/or release of these species in the gas phase has increased temperatures, but we are unable to confirm if the calculated excitation temperatures ($T_\mathrm{ex} \sim 10-30$\,K) are truly enhanced due to the limitations of our low signal-to-noise lines.
    
    \item 
    We find that COM abundances with respect to CH$_3$OH for IRAS\,16293E, A, and\,B are similar within factors of $<3$ (with the exception of CH$_3$CHO for the `A' source). Because these three sources have all formed from the same natal cloud environment, we show directly for this system in particular that the relative abundances of COMs and D-COMs in the later protostellar stage were set during the earlier prestellar stage. 
    
    \item The D/H ratios (with statistical correction) for CH$_2$DOH are similar within errors, but increase slightly from prestellar cores to protostars, while, in contrast for CHD$_2$OH, they decrease slightly. More specifically, from sources IRAS\,16293E, B, and A these CHD$_2$OH/CH$_2$DOH ratios are 35$\pm15\%$, 25$\pm5.0\%$ and 20$\pm6.8\%$, respectively.
    The D/H enhancement in doubly deuterated methanol is seen also across a heterogeneous sample of starless and prestellar cores, as well as a select number of protostars and the comet 67P/C-G, further supporting a chemical inheritance from the prestellar stage.
    In the context of these other sources, abundance comparisons of the higher complexity COMs targeted in this study reveal that IRAS\,16293E is a `typically complex prestellar core'. Though, the influence of the nearby outflows and shocked gas could enhance gas phase abundances of other COMs not searched for thus far. 
    
\end{enumerate} 

Future work to characterize the spatial distribution and solid phase inventory (e.g., ice compositions using JWST) of COMs and D-COMs in and around IRAS\,16293E and its natal cloud will help shed more light on formation pathways and on the influence of cloud environments (e.g., shocks, outflows, irradiation) on the chemistry itself.  

\begin{acknowledgements}
{We thank the anonymous referee for their feedback, which significantly improved the manuscript.} We are thankful that we have the opportunity to conduct astronomical research on Iolkam Du'ag (Kitt Peak) in Arizona and we recognize and acknowledge the very significant cultural role and reverence that these sites have to the Tohono O'odham Nation.
We sincerely thank the operators of the Arizona Radio Observatory for their assistance with the observations. 
The 12\,m Telescope is operated by the Arizona Radio Observatory (ARO), Steward Observatory, University of Arizona, with funding from the State of Arizona, NSF MRI Grant AST-1531366 (PI Ziurys), NSF MSIP grant SV5-85009/AST- 1440254 (PI Marrone), NSF CAREER grant AST-1653228 (PI Marrone), and a PIRE grant OISE-1743747 (PI Psaltis). 

We also acknowledge observations carried out with the Yebes 40 m telescope (24A006). Paula Tarr\'io carried out the observations and the first inspection of the data quality. The 40\,m radio telescope at Yebes Observatory is operated by the Spanish Geographic Institute (IGN; Ministerio de Transportes, Movilidad y Agenda Urbana). Yebes Observatory thanks the European Union’s Horizon 2020 research and innovation programme for funding support to ORP project under grant agreement No 101004719. 

S.S. acknowledges the National Radio Astronomy Observatory is a facility of the National Science Foundation operated under cooperative agreement by Associated Universities, Inc. M.N.D. acknowledges the Holcim Foundation Stipend. B.M.K acknowledges the SNSF Postdoc.Mobility stipend P500PT\_214459. J. Ferrer Asensio thanks RIKEN Special Postdoctoral Researcher Program (Fellowships) for financial support. Y.S. acknowledges the National Science foundation Astronomy and Astrophysics grant AST-2205474.
\end{acknowledgements}

%
%

\bibliographystyle{aa} 
\bibliography{references.bib}

\begin{appendix}

\onecolumn
\begin{landscape}
\section{Supplemental table}

\begin{table*}[h!]
\caption{ Abundance comparisons across cores} 
\label{tab:compare_cores}      
\setlength{\tabcolsep}{3pt}
\centering                          
\begin{tabular}{c c c c c c c c c c c c}       
\hline\hline          \\       
Name & Class  & Telescope & $\frac{^{13}\mathrm{CH}_3\mathrm{OH}}{\mathrm{CH}_3\mathrm{OH}}$ $\times10^{-2}$ & $\frac{\mathrm{CH_2DOH}}{\mathrm{CH}_3\mathrm{OH}}$ $\times10^{-2}$& $\frac{\mathrm{CHD_2OH}}{\mathrm{CH}_3\mathrm{OH}}$ $\times10^{-2}$&$\frac{\mathrm{CH_3CHO}}{\mathrm{CH}_3\mathrm{OH}}$ $\times10^{-2}$&   $\frac{\mathrm{HCOOCH_3}}{\mathrm{CH}_3\mathrm{OH}}$ $\times10^{-2}$& $\frac{\mathrm{CH_3OCH_3}}{\mathrm{CH}_3\mathrm{OH}}$ $\times10^{-2}$& $\frac{\mathrm{CH_2CHCN}}{\mathrm{CH}_3\mathrm{OH}}$ $\times10^{-2}$ & Ref. \\ [5pt] 
TMC-1  & starless                     & SD      & 1.5$\pm$0.4 & --          & --          & 9.7$\pm$1.5    & 3.4$\pm$0.7  & 7.8$\pm$2.3 & 20$\pm$5.3  & (2,16,30,31)\\  
L1521E & starless                     & SD      & --          & 0.6$\pm$0.5 & $<0.7$      & 13$\pm$5       & 47$\pm$20    & 10$\pm$2.5  & 4.0$\pm$1.6 & (24,27,33)\\  
B68    & starless                     & SD      & --          & $<1.4$      & $<2.8$      &  --            &     --       &   --        &   --        & (24)\\  
L1495-B10\tablefootmark{a}&starless \& prestellar& SD      & --          & 9.8$\pm$4.8 & --           & --             & --           & --          & --          & (3) \\ 
L1495\tablefootmark{a}&starless \& prestellar    & SD      & --          & --          & --           & 9.2$\pm$2.1    & --           & --          & --          & (32) \\ 
Perseus\tablefootmark{a} &starless \& prestellar & SD      & --          & --          & --           & 5.0$\pm$0.2    & 4.5$\pm$0.3  & 7.7$\pm$3.4 & 0.2$\pm$0.4 & (34) \\ 
L183 & prestellar                     & SD      & --          & 3.9$\pm$0.1 & --           & 5.5$\pm$1.8    & 2.4$\pm$0.4  & --          & --          & (23) \\ 
L1689B & prestellar                   & SD      & --          & --          & --           & 14$\pm$11      & 30$\pm$24    & 11$\pm$11   & --          & (6) \\ 
L694-2 & prestellar                   & SD      & --          & 2.5$\pm$1.5 & 1.3$\pm$0.8 & --             & --           & --          & --          & (24) \\ 
H-MM1  & prestellar                   & SD      & --          & 5.6$\pm$2.5 & 4.4$\pm$1.6 & --             & --           & --          & --          & (24) \\ 
L1544  & prestellar                   & SD      & --          & 7.2$\pm$2.0 & --          & 1.9$\pm$0.5    &  6.7$\pm$6.4 & 2.3$\pm$0.7 & 1.8$\pm$1.3 & (13,18,20,24) \\ 
IRAS\,16293E &  prestellar           & SD      & 1.4$\pm$0.2 & {6.2$\pm$1.8} & {2.2$\pm$0.7} & {3.0$\pm$1.0}   & {4.0$\pm$0.1} & {2.1$\pm$0.6}  & $<$0.04     &\textit{this work} \\ 
IRAS\,16293A & protostar (Class 0)  & Int     &1.5$\pm$0.7  & 8.5$\pm$3.5 & 1.7$\pm$0.6 &  0.03$\pm$0.01 & 2.1$\pm$0.9  & 4.0$\pm$1.7 & 0.002$\pm$0.001 & (12, 15, 26) \\  
IRAS\,16293B & protostar (Class 0)  & Int     &1.2$\pm$0.5  & 7.1$\pm$1.4 & 1.8$\pm$0.5 &  1.2$\pm$0.24  & 2.6$\pm$0.5  & 2.4$\pm$0.5 & 0.007$\pm$0.002 & (12, 15, 19) \\  
               &                      & Int     &1.5$\pm$0.3  & 1.8$\pm$0.3 & 2.3$\pm$0.6 &  0.4$\pm$0.1   & 1.2$\pm$0.2  & 1.0$\pm$0.4 & $<$0.05         & (28) \\  
IRAS\,16293    & protostar (Class 0)  & SD      & --          & {43$\pm$8.6} 
& 17$\pm$6    &  17$\pm$21     & 133$\pm$35   & 79$\pm$123  & --              & (10,11)\\  
IRAS\,2A & protostar (Class 0)        & SD      & --          & {26$\pm$10}    
&  21$\pm$8.6 &  --            &  $<$85       & $<$53       & --              & (5,25,29) \\ 
         &                            & Int     & --          & 5.8$\pm$4.1 & 4.4$\pm$3.1 &  --            &  1.9$\pm$0.8 & 1.2$\pm$0.6 & --              & (35,36) \\
IRAS\,4A & protostar (Class 0)        & SD      & --          & {31$\pm$7.0}  
& 16$\pm$4.1  &  --            &  52          & $<$22       & --              & (4,25,29)\\ 
         &                            & Int     & --          & 3.7$\pm$2.3 & 2.1$\pm$1.1 &  --            &  1.5$\pm$1.0 & 0.9$\pm$0.4 & --              & (35,36)\\ 
IRAS\,4B & protostar (Class 0)        & SD      & --          & {18$\pm$6.2}  
& 11$\pm$4.0  &  13            &  $<$19       & --          &--               & (5,29) \\
L483 & protostar (Class 0)            & SD      &  --         & 1.9$\pm$0.4 & 0.3$\pm$0.1 & 1.3$\pm$0.3    & 0.8$\pm$0.2  & 1.8$\pm$0.5 &  --             & (1) \\ 
HH212 & protostar (Class 0)           & Int     & 1.9$\pm$1.5 & 2.4$\pm$0.4 & --          & 2.4$\pm$0.6    & 5.3$\pm$1.3  & --          & --              & (8,22) \\ 
L1157-B1 & protostar (outflow)        & SD      & --          &  --         & --          & 2.8$\pm$0.5    & 5.4$\pm$1.0  & 4.9$\pm$0.9 &  --             & (21) \\ 
SVS13A & protostar (Class I)          & SD      & --          & 4.0$\pm$2.8 & --          & 0.17$\pm$0.03  & 1.9$\pm$0.4  & 1.8$\pm$0.4 &  --             & (7) \\ 
67P/C-G & comet                       & Rosetta &--           &\multicolumn{2}{c}{0.7-6.6}& 150            &  44          &  3          & --              &  (14,17) \\ 
\hline    
\end{tabular}
     \tablefoot{Telescope denoted as `SD' for single dish and 'Int' for Interferometer. Note: the ratios involving deuterated species represent column density ratios, not D/H statistical ratios. \tablefoottext{a}{Represents average within a larger sample of cores.} References (Ref.): 
     (1) \cite{2019A&A...625A.147A} 
     (2) \cite{2021A&A...649L...4A} 
     (3) \cite{2021MNRAS.501..347A} 
     (4) \cite{2004ApJ...615..354B} 
     (5) \cite{2007A&A...463..601B} 
     (6) \citealt{2012A&A...541L..12B} 
     (7) \cite{2017MNRAS.467.3011B} 
     (8) \cite{2017A&A...606L...7B} 
     (9) \cite{2019MNRAS.483.1850B} 
     (10) \cite{2002A&A...393L..49P} 
     (11) \cite{2003ApJ...593L..51C} 
     (12) \cite{2018A&A...616A..90C} 
     (13) \cite{2019A&A...622A.141C}   
     (14) \cite{2021MNRAS.500.4901D} 
     (15) \cite{2022A&A...659A..69D} 
     (16) \cite{2016ApJS..225...25G}   
     (17) \citep{2023A&A...678A..22H} 
     (18) \cite{2016ApJ...830L...6J}  
     (19) \cite{2018A&A...620A.170J} 
     (20) \cite{2021ApJ...917...44J} 
     (21) \cite{2017MNRAS.469L..73L}  
     (22) \cite{2019ApJ...876...63L} 
     (23) \cite{2020A&A...633A.118L} 
     (24) \cite{2023A&A...669L...6L} 
     (25) \cite{2005A&A...442..527M} 
     (26) \cite{2020A&A...635A..48M} 
     (27) \cite{2019A&A...630A.136N} 
     (28) \cite{2024A&A...686A..59N} 
     (29) \cite{2006A&A...453..949P} 
     (30) \cite{2015ApJ...802...74S} 
     (31) \cite{2018ApJ...854..116S} 
     (32) \cite{2020ApJ...891...73S} 
     (33) \cite{2021MNRAS.504.5754S} 
     (34) \cite{2024MNRAS.533.4104S} 
     (35) \cite{2015ApJ...804...81T} 
     (36) \cite{2019A&A...632A..19T} 
     (37) \cite{2014ApJ...795L...2V}. 
     Note that the CH$_2$DOH values reported in (10) and (29) have been overestimated by a factor of 2 (see footnote 7 in \citealt{2016A&A...587A..91B}) and we correct for this factor in our estimates. }
\end{table*}
\end{landscape}

\twocolumn
\section{RADEX minimization}
\label{appendix_radex_min}

\begin{figure}[tbh]
\begin{center}$
\begin{array}{c}
\includegraphics[width=85mm]{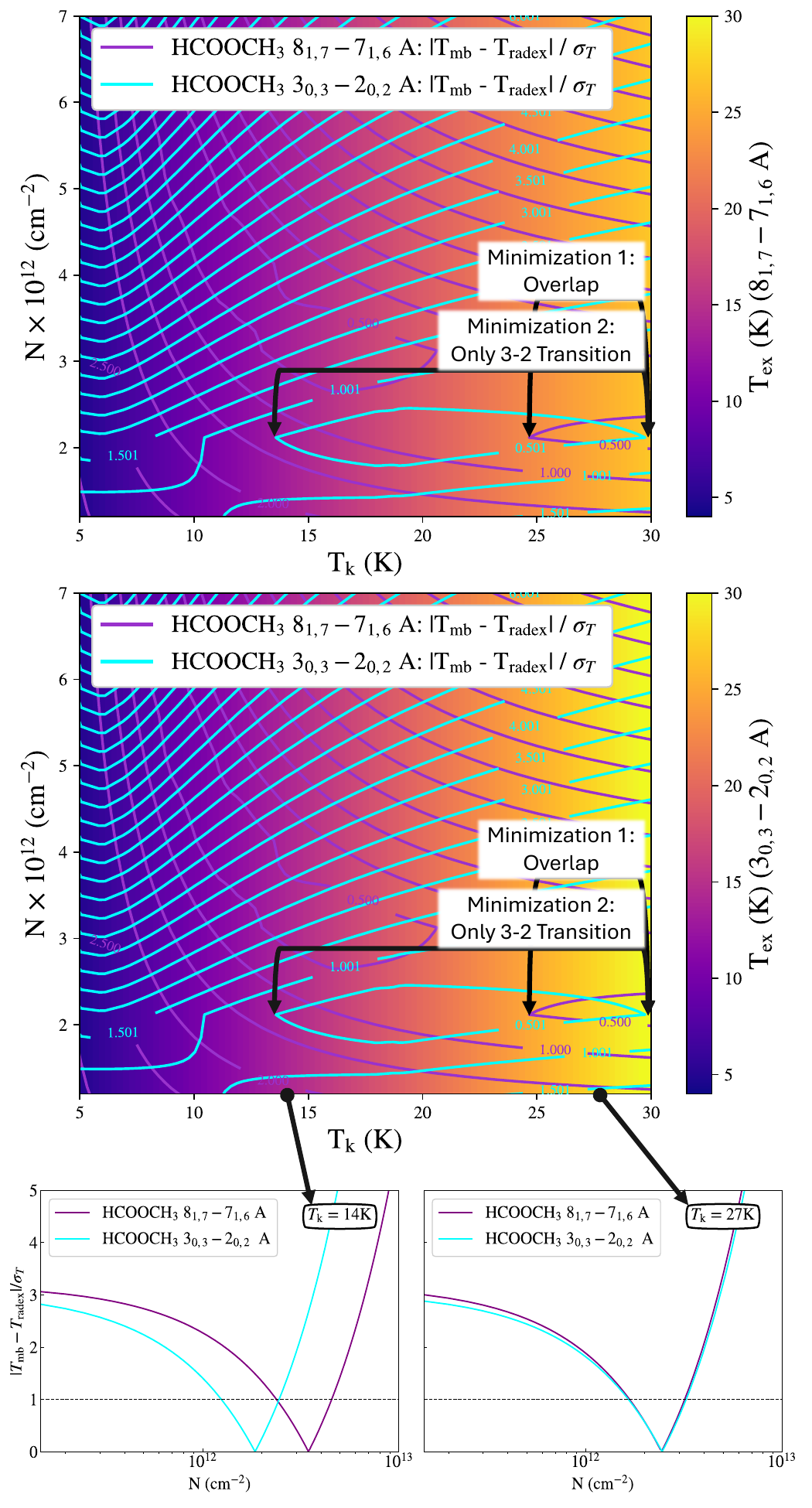}
\end{array}$
\end{center}
\caption{ Grids of column density, $N$, versus kinetic temperature, $T_\mathrm{k}$, for the case of HCOOCH$_3$ A when excitation temperature, $T_\mathrm{ex}$, is calculated at each point for the $8_{1,7}- 7_{1,6}$ A transition (top) and the $3_{0,3}-2_{0,2}$ A transition (middle). Overlaid on the grids are minimization contours for these two `A' state transitions, where $8_{1,7}- 7_{1,6}$ A is in purple and $3_{0,3}-2_{0,2}$ A is in cyan. The volume density is set to $7.72\times 10^5$\,cm$^{-3}$ and $9.75\times 10^5$\,cm$^{-3}$ for the $8_{1,7}- 7_{1,6}$ A and $3_{0,3}-2_{0,2}$ A lines, respectively. In the case of `Minimization 1', black arrows denote when the two modeled transitions are both best matched to the observations and overlap in $N$ and $T_\mathrm{k}$ space.  In the case of `Minimization 2', black arrows denote where in parameter space we would expect to be if only the $3_{0,3}-2_{0,2}$ transition is considered. Minimization curves at the bottom of the figure are also extracted at slices of the grid where $T_\mathrm{k} = 14$\,K, and $T_\mathrm{k} = 27$\,K to further illustrate how a best-fit is determined.
}
\label{fig:RADEX_grid_HCOOCH3}
\end{figure}

Here, we show supplemental plots for the iterative non-LTE radiative transfer RADEX fitting procedure described in Section\,\ref{sec:subRADEX}. Grids of input column density, $N$, at a given kinetic temperature, $T_\mathrm{k}$, for each transition detected are run and then minimized by subtracting the observed peak line temperature, $T_\mathrm{{mb}}$, from the RADEX-derived peak line temperature, $T_\mathrm{{radex}}$, and dividing it by the observed noise, $\sigma_T$. When this prescription, $|T_\mathrm{{mb}} - T_\mathrm{{radex}}| / \sigma_T$, is below a value of `1' this signifies a `best-fit'. To illustrate this parameter space, in Figure\,\ref{fig:RADEX_grid_HCOOCH3} we plot the grid of `N' versus $T_\mathrm{k}$ for the case of of HCOOCH$_3$ where the color-scale represents $T_\mathrm{ex}$ for a specific transition. Overlaid on the grid are minimization contours which show where the best fits of both `A' state transitions overlap. Subplots are also shown to illustrate what those minimization curves look like at specific $T_\mathrm{k}$ slices in the grid. For HCOOCH$_3$ A, we only have these two transitions to minimize over, so we assume $f=1$ (when the source size is much bigger than the beam) and fix the volume density inputs (to $7.72\times 10^5$\,cm$^{-3}$ and $9.75\times 10^5$\,cm$^{-3}$ for the $8_{1,7}- 7_{1,6}$ A and $3_{0,3}-2_{0,2}$ A lines, respectively) based on beam-averaged \textit{Herschel} observations (described in section\,\ref{sec:subRADEX}). It should be noted, however, that for our molecules of interest the best-fit column density, $N$, is rather insensitive to the the beam-averaged volume density, $n_\mathrm{beam}$ across a range of $T_\mathrm{k}$ values (see Appendix\,\ref{appendix_radex_uncertain}).

As described in section\,\ref{sec:subRADEX}, the 100\,GHz HCOOCH$_3$ $8_{1,7} - 7_{1,6}$ A line should be treated with caution because it has been smoothed by three channels to increase the signal-to-noise ratio and it has a linewidth significantly ($\sim 3\times$) broader than the corresponding 36\,GHz $3_{0,3} - 2_{0,2}$ A line (Table\,\ref{tab:lines}). By treating the $3\sigma$ detected $8_{1,7} - 7_{1,6}$ A transition as an upper limit, our best-fit column density constraint remains robust (at $\sim2 \times 10^{12}$\,cm$^{-2}$) but the best-fit kinetic temperature range can vary by more than a factor of two, from $T_\mathrm{k} \sim14-30$\,K (Figure\,\ref{fig:RADEX_grid_HCOOCH3}). It should also be noted that the excitation temperature at each point on the $N$ and $T_\mathrm{k}$ grid varies by $\sim5-10$\,K depending on which transitions is plotted, i.e., the $T_\mathrm{ex}$ for the $8_{1,7}- 7_{1,6}$ A transition is $\sim$20\,K and for the $3_{0,3}-2_{0,2}$ A transition it is $\sim30$\,K at a $T_\mathrm{k} = 27\,K$ (compare color-scales in the two grids in Figure\,\ref{fig:RADEX_grid_HCOOCH3}).  

\begin{figure*}[h!]
\begin{center}$
\begin{array}{c}
\includegraphics[width=\textwidth]{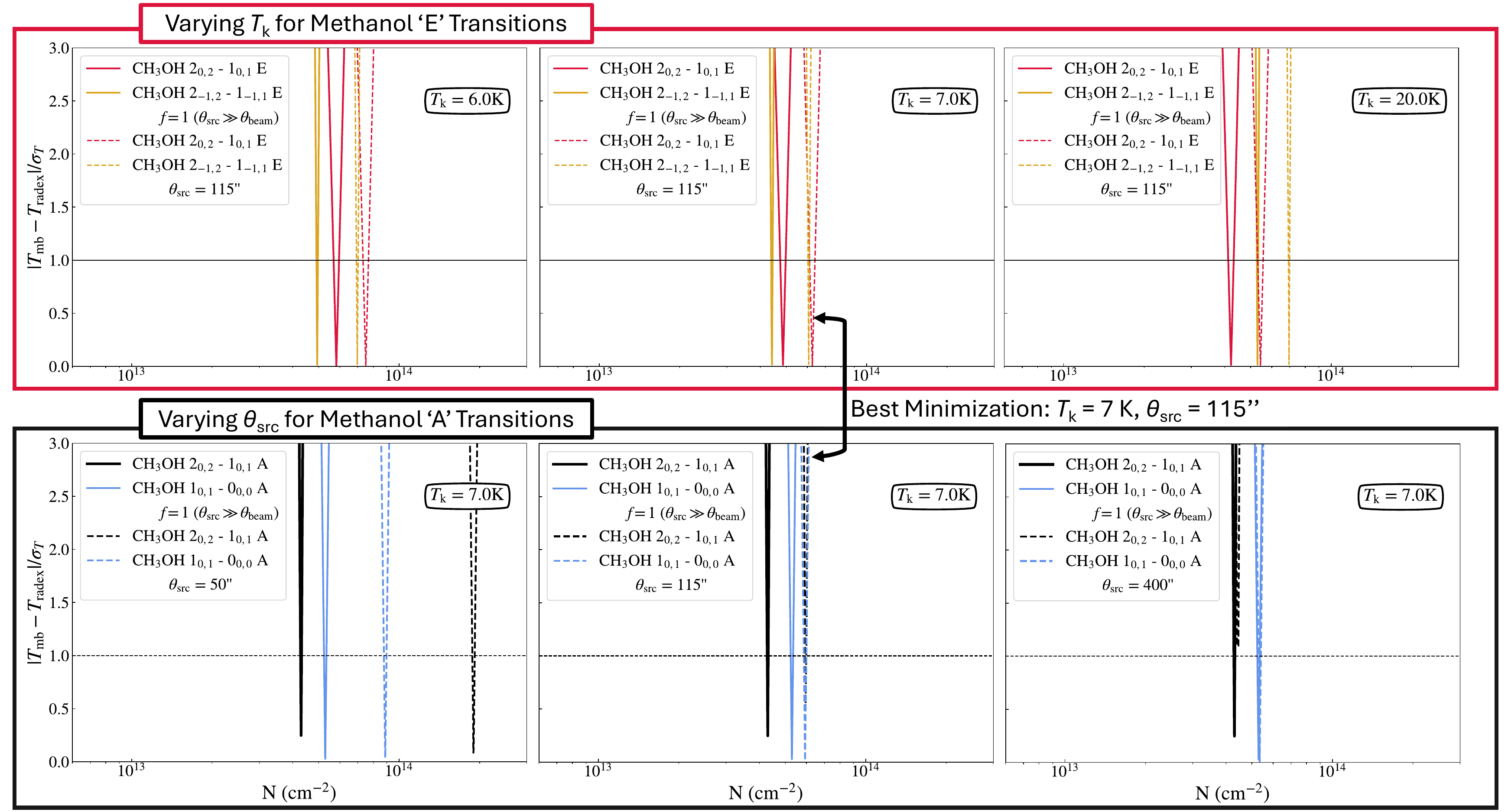}
\end{array}$
\end{center}
\caption{ Representative RADEX CH$_3$OH minimization plots for IRAS\,16293E (see also Figure C1 in \citealt{2024MNRAS.533.4104S}). Plotted in each panel is the difference in the observed brightness temperature, $T_\mathrm{mb}$, and the RADEX-calculated brightness temperature, $T_\mathrm{radex}$, divided by the noise level, $\sigma_T$, versus the RADEX-calculated column density, $N$. When the RADEX model best fits the observations, $| T_\mathrm{mb} - T_\mathrm{radex}|/\sigma_T$ is minimized and the best-fit $N$ is found. In the top panels minimization curves are shown for the CH$_3$OH E transitions from the ARO 12 m observations at different $T_\mathrm{k}$ values (in upper right corner of each plot). In the bottom panels minimization curves are shown for CH$_3$OH A transitions, from both the ARO 12 m and Yebes 40 m observations, at a fixed $T_\mathrm{k} = 7$\,K and for varying $\theta_\mathrm{src}$ values. In each plot the solid lines indicate when $f = 1$ is assumed and the dashed lines indicate that a specific $\theta_\mathrm{src}$ has been used in the fit.}
\label{fig:sourcesize}
\end{figure*}

The RADEX parameter space increases when more than two transitions are modeled and when the source size, $\theta_\mathrm{src}$, is allowed to vary, making the minimization multi-dimensional and more challenging to visualize. Therefore, in the case of CH$_3$OH we show representative slices of the minimization grid to illustrate how both $T_\mathrm{k}$ and $\theta_\mathrm{src}$ were constrained (Figure\,\ref{fig:sourcesize}). For this molecule, we have multiple transitions observed with different telescope beam sizes, $\theta_\mathrm{beam}$, and we can use this information to solve for $\theta_\mathrm{src}$ in equation\,\ref{fillingfrac} when the minimized column density of one transition from one telescope (i.e., CH$_3$OH 2$_{0,2}$ - 1$_{0,1}$ A from the ARO 12 m) best matches the column density of another transition from the other telescope (i.e., CH$_3$OH 1$_{0,1}$ - 0$_{0,0}$ A from the Yebes 40 m). However, before we can do that, we need to first constrain $T_\mathrm{k}$, since this is our other free parameter in the fit aside from $N$. 
This is done with the CH$_3$OH E transitions that both come from the ARO 12m (same $\theta_\mathrm{beam}$) so as to better control for any source size variation. 

We find that {the two CH$_3$OH E minimization curves (2$_{0,2}$ - 1$_{0,1}$ E and 2$_{-1,2}$ - 1$_{-1,1}$ E), plotted for a standard $f=1$ (solid lines) in the top row of Figure\,\ref{fig:sourcesize}, overlap most closely at a $T_\mathrm{k}$ of $7$\,K.}
At a fixed $T_\mathrm{k}$ of 7\,K, we can now minimize over $\theta_\mathrm{src}$ with the CH$_3$OH A transitions (2$_{0,2}$ - 1$_{0,1}$ A and 1$_{0,1}$ - 0$_{0,0}$ A observed with the ARO 12\,m and Yebes 40\,m, respectively), plotted as dashed lines at representative values of 50\,\arcsec, 115\,\arcsec, and 400\,\arcsec in the bottom row of Figure\,\ref{fig:sourcesize}. The best minimization where both `A' transitions overlap at a $T_\mathrm{k}$ of 7\,K is when $\theta_\mathrm{src}$ = 115\,\arcsec. In this iterative fitting, we then go back and plot updated minimization curves for the `E' transitions for $\theta_\mathrm{src}$ = 115\,\arcsec (dashed lines in top panels of Figure\,\ref{fig:sourcesize}) in order to extract the best-fit $N$ values (see Table\,\ref{table:colden} for the exact values from the full fitting procedure). 
We note that this procedure does assume that $T_\mathrm{k}$ and $\theta_\mathrm{src}$ for both the A and E states are the same. 

\section{RADEX sources of uncertainty} \label{appendix_radex_uncertain}

\begin{figure*}[h!]
\begin{center}$
\begin{array}{cc}
\includegraphics[width=85mm]{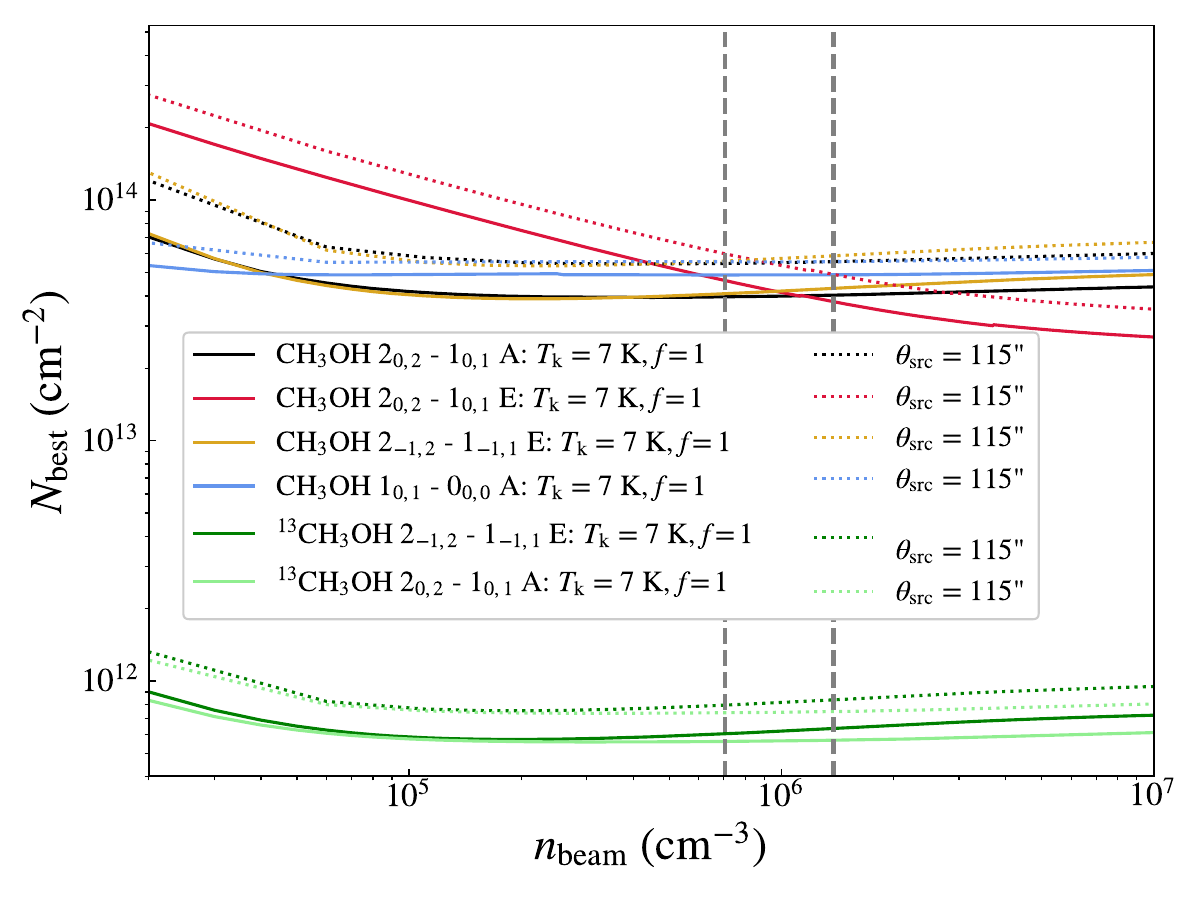}& 
\includegraphics[width=85mm]{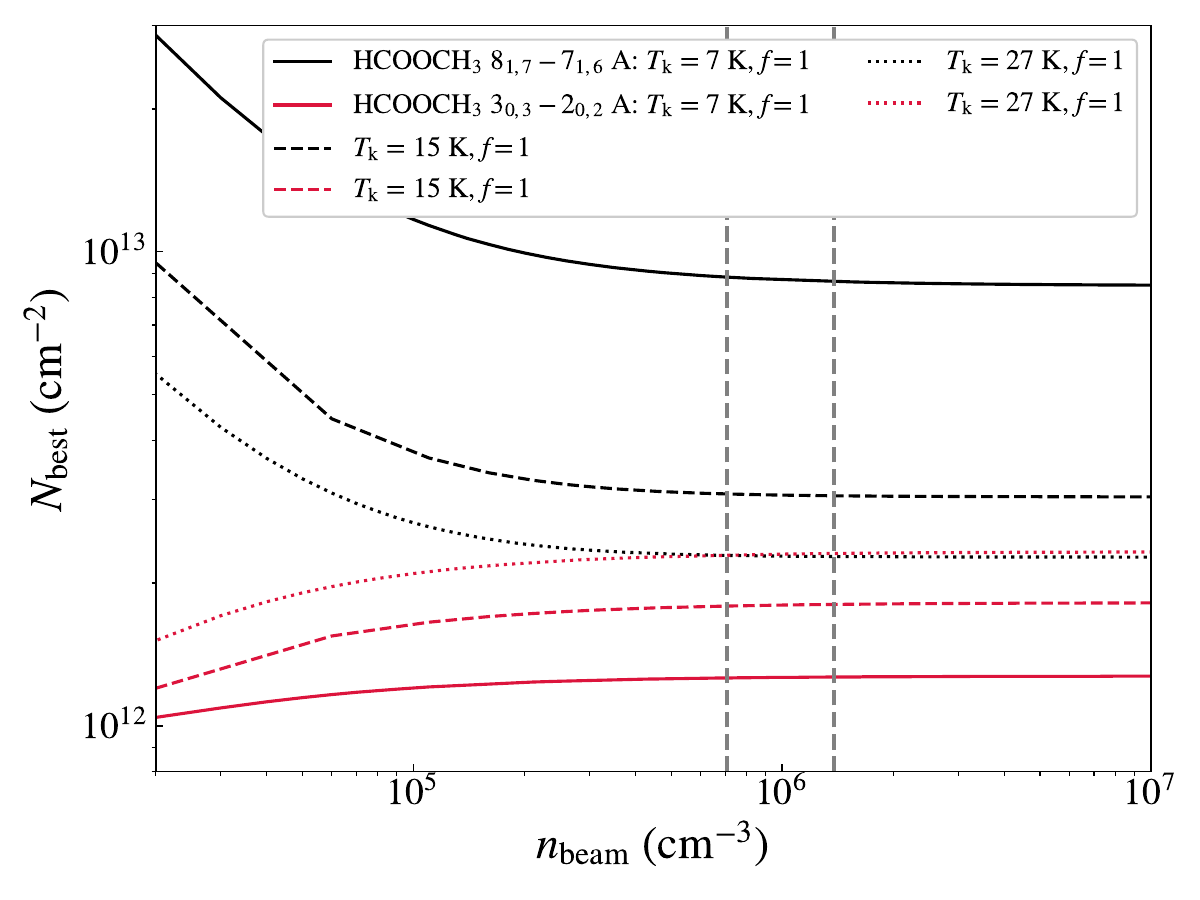}
\end{array}$
\end{center}
\caption{ {RADEX-calculated} column density, $N_\mathrm{best}$, variation for different inputs of the beam-average volume density, $n_\mathrm{beam}$. { (left) Methanol (CH$_3$OH and $^{13}$CH$_3$OH) transitions are plotted in the case assuming a filling fraction of 1 ($f=1$) or $\theta_\mathrm{src}=115$\,\arcsec. (right) Methyl formate (HCOOCH$_3$ A) transitions are plotted for different kinetic temperature, $T_\mathrm{k}$, values.} The vertical gray dashed lines show the range in beam-averaged volume densities used in our analysis.  }
\label{fig:N_vs_n}
\end{figure*}

One of the main sources of uncertainty in RADEX calculations can be the averaged volume density, $n_\mathrm{beam}$, which, depending on the molecule, may affect the RADEX-calculated column density. To explore the dependence of the derived column density on the input $n_\mathrm{beam}$ we run new grids. This time, we let RADEX solve for the column density, which we label here as $N_\mathrm{best}$, given our directly observed inputs (e.g., the measured $T_\mathrm{mb}$) while varying $n_\mathrm{beam}$ from $10^4 - 10^7$\,cm$^{-3}$. 

For the molecules {modeled with RADEX}, CH$_3$OH, $^{13}$CH$_3$OH, and HCOOCH$_3$, we find that $N_\mathrm{best}$ is rather insensitive to the choice of $n_\mathrm{beam}$ at values roughly $>10^5$\,cm$^{-3}$ (Figure\,\ref{fig:N_vs_n}). This same dependence was also found for the case of CH$_3$OH in \cite{2020ApJ...891...73S} (see their Figure 6). And, within the total range of $n_\mathrm{beam}$ values calculated for each transition, namely 0.71 - 1.38 $\times 10^6$\,cm$^{-3}$ (described in Section\,\ref{sec:subRADEX}), $N_\mathrm{best}$ is flat with a difference of at most a factor of 1.2, which is the case only for CH$_3$OH 2$_{0,2}$ - 1$_{0,1}$ E (note: dashed gray lines in Figure\,\ref{fig:N_vs_n} show this range in $n_\mathrm{beam}$). Additionally, we find that changing $\theta_\mathrm{src}$ does not alter the dependence on $n_\mathrm{beam}$ (left panel of Figure\,\ref{fig:N_vs_n}).

For HCOOCH$_3$, we demonstrate how {the $N_\mathrm{best}$ versus $n_\mathrm{beam}$ curves scale with the input kinetic temperature, $T_\mathrm{k}$. For the two `A' state transitions ($8_{1,7}- 7_{1,6}$ A in black and $3_{0,3}-2_{0,2}$ A in red) representative curves for $T_\mathrm{k}= 7$\,K (solid line), $T_\mathrm{k} = 15$\,K (dashed line), and $T_\mathrm{k} = 27$\,K (dotted line) are plotted in the right panel of Figure\,\ref{fig:N_vs_n}. We find changing $T_\mathrm{k}$ does not alter the dependence on $n_\mathrm{beam}$. As also seen in our minimization grid (Figure\,\ref{fig:RADEX_grid_HCOOCH3}), and again assuming here that $\theta_\mathrm{src} \gg \theta_\mathrm{beam}$ or $f=1$, the $N_\mathrm{best}$ for both transitions converges at a $T_\mathrm{k}$ of 27\,K, i.e., where the dotted red curve meets {the} dotted black curve (right panel of Figure\,\ref{fig:N_vs_n}).}

\section{Supplemental LTE fitting}\label{appendix_LTE}

\begin{figure}[tbh]
\begin{center}$
\begin{array}{cc}
\includegraphics[width=75mm]{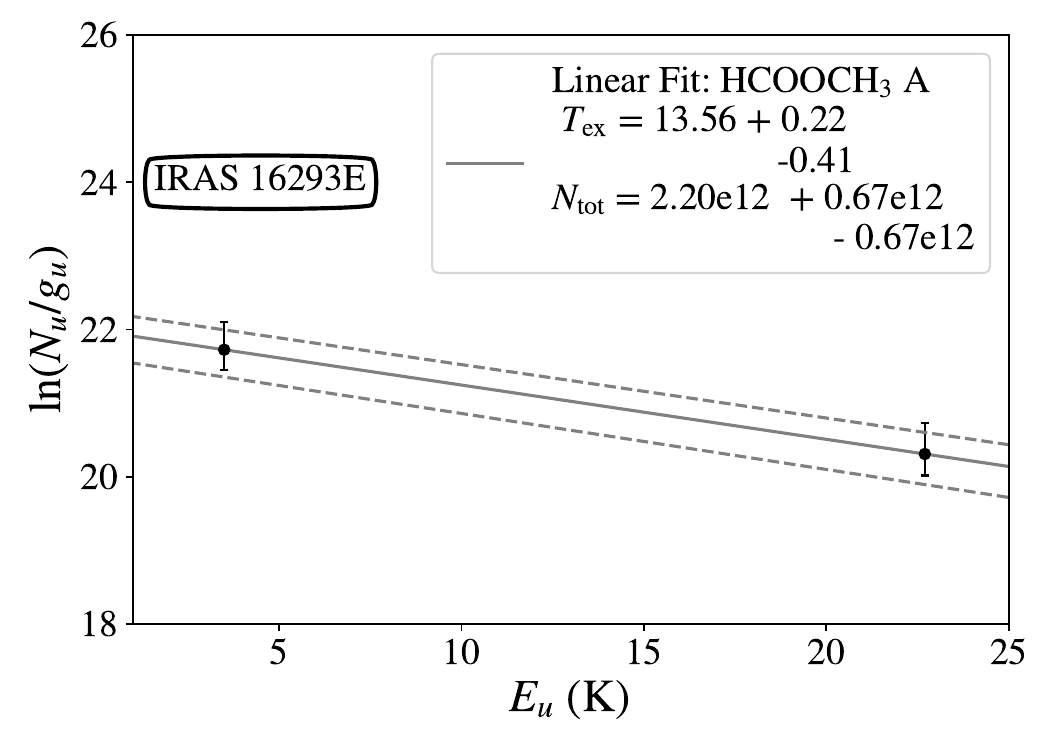}
\end{array}$
\end{center}
\caption{ Rotation diagram with associated linear best-fits (solid curves) and corresponding uncertainty (dashed curves) for HCOOCH$_3$, using the two A state transitions detected towards IRAS 16293E. 
}
\label{fig:LTE_MF}
\end{figure}

We explore different LTE methods as supplemental analysis for the COMs HCOOCH$_3$ and CH$_3$OCH$_3$. In Figure\,\ref{fig:LTE_MF}, we plot a rotation diagram (method described in Section\,\ref{sec:subLTE}) for HCOOCH$_3$ A and find $T_\mathrm{ex} = 13.56^{+0.22}_{-0.41}$\,K and $N = 2.2^{+0.67}_{-0.67} \times 10^{12}$\,cm$^{-2}$. The fitted column density is consistent with the RADEX minimization method, but $T_\mathrm{ex}$ is a factor of two lower (see Table\,\ref{table:colden}). We know from Figure\,\ref{fig:RADEX_grid_HCOOCH3} in Appendix\,\ref{appendix_radex_min} that this value is still within RADEX's `best-fit' parameter space and indicates $T_\mathrm{ex}$ derived from the LTE approach is at the lower end of the range. 
We therefore stress that due to our limited number of transitions at low signal-to-noise {ratios} and possible non-LTE effects (i.e., in the RADEX calculations $T_\mathrm{ex}$ for the two separate transitions can vary by $\sim5-10\,K$ for the same $T_\mathrm{k}$; see Figure\,\ref{fig:RADEX_grid_HCOOCH3}), the exact $T_\mathrm{ex}$ for HCOOCH$_3$ in IRAS 16293E is still uncertain.

\begin{figure}[tbh]
\begin{center}$
\begin{array}{cc}
\includegraphics[width=75mm]{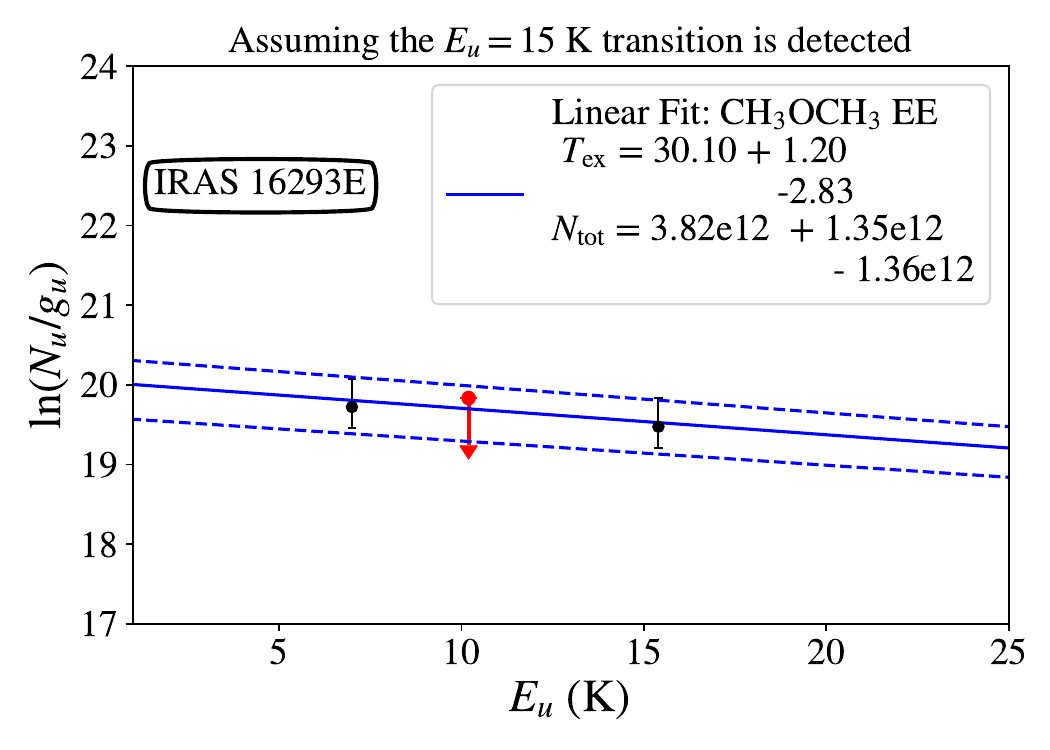}
\\
\includegraphics[width=75mm]{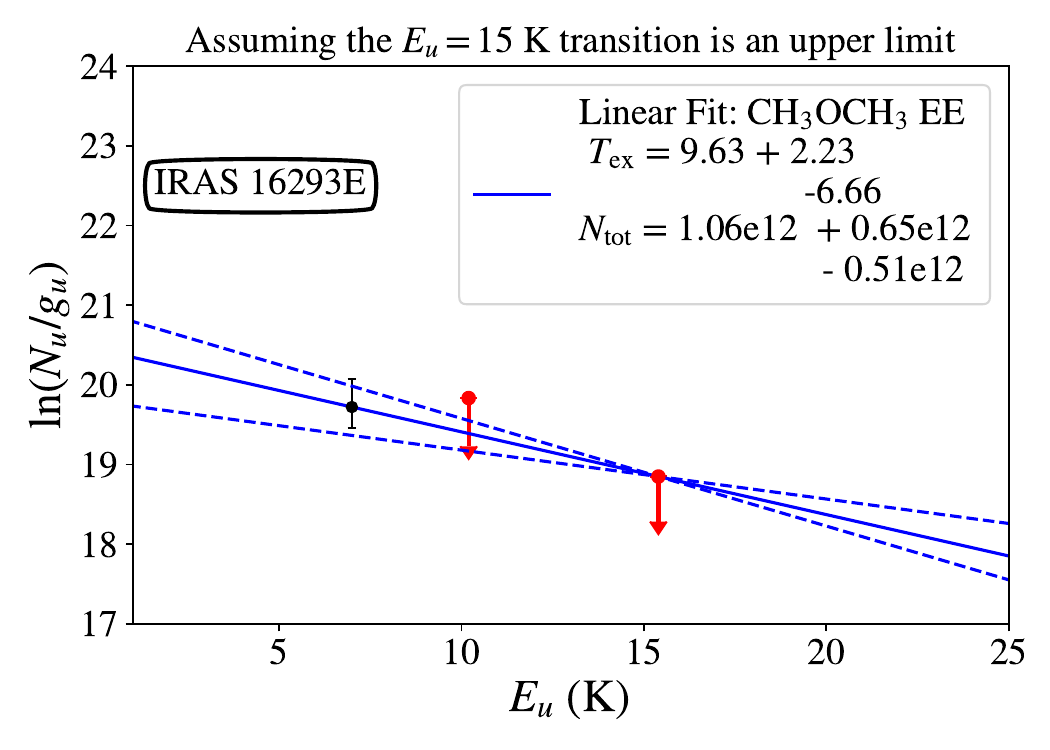}
\end{array}$
\end{center}
\caption{ Rotation diagram with associated linear best-fits (solid curves) and corresponding uncertainty (dashed curves) for CH$_3$OCH$_3$ using the EE state transitions observed toward IRAS 16293E assuming either the $5_{1,4}-5_{0,5}$ EE transition ($E_\mathrm{u}$=15.4\,K) is a detection (top) or upper limit (bottom). 
}
\label{fig:dme_rot}
\end{figure}

\begin{figure}[tbh]
\begin{center}$
\begin{array}{cc}
\includegraphics[width=85mm]{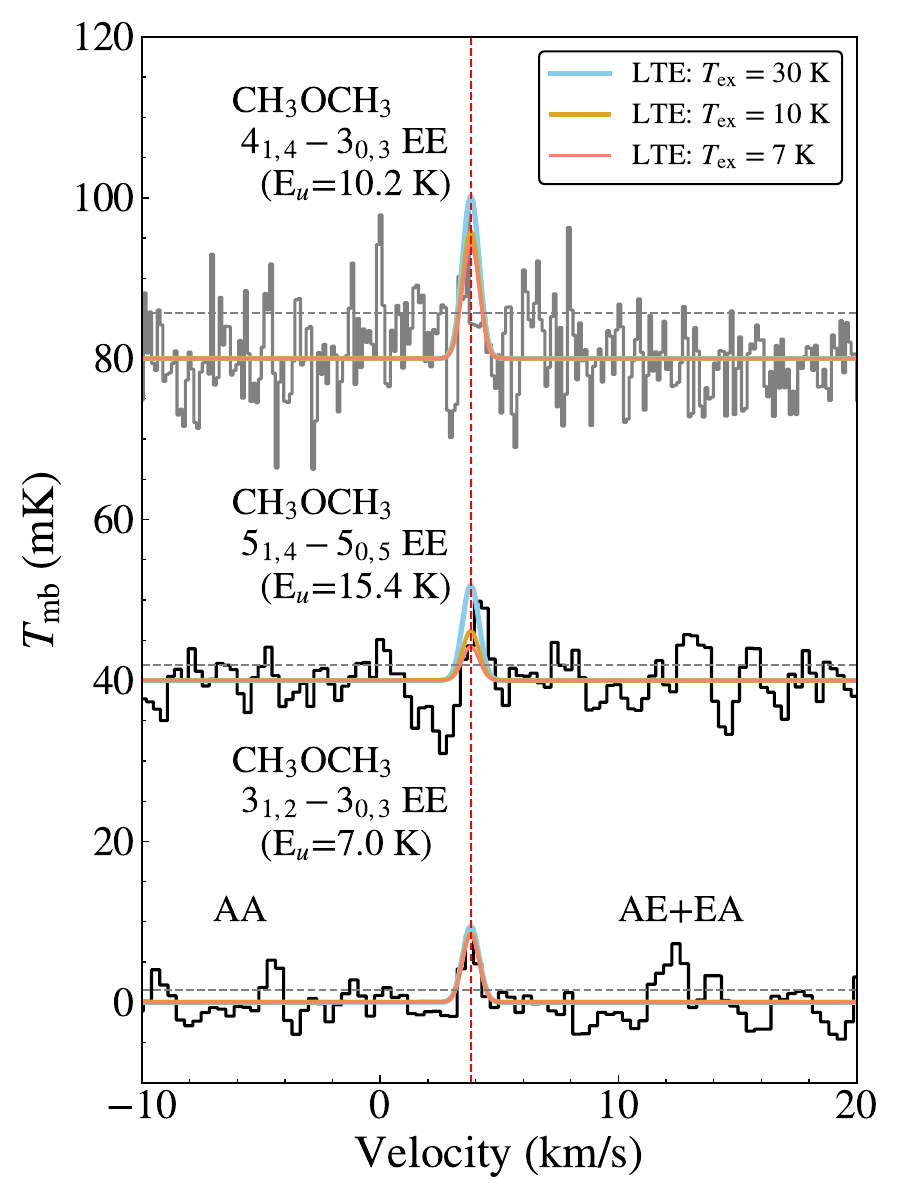}
\end{array}$
\end{center}
\caption{ {Synthetic spectra fits to the CH$_3$OCH$_3$ EE lines from the ARO 12 m (top) and Yebes 40 m (middle and bottom). Note that the ARO 12 m transition (the $4_{1,4} - 3_{0,3}$ EE line) was not detected above $3\sigma$ and therefore only provides an upper limit. Spectra are offset by {40}\,mK for easier viewing. The vertical red line is centered at a $v_\mathrm{lsr}$ of 3.8\,km/s { and the gray horizontal lines show the $1\sigma$ noise ($rms$) level. Only for the $3_{1,2}-3_{0,3}$ transition are the AA and AE+EA torsional states detected above the noise.}
}
}
\label{fig:LTE_SYN_DME}
\end{figure}

In the case of CH$_3$OCH$_3$ we plot two separate rotation diagrams for the `EE' states, one if the $5_{1,4}-5_{0,5}$ EE transition ($E_\mathrm{u}$=15.4\,K) is considered a detection {($T_\mathrm{mb}$ = 10.6\,mK; Table\,\ref{tab:lines})} and one of it is considered an upper limit ({$T_\mathrm{mb}$ = 3$\sigma$ = 5.7\,mK}; Figure\,\ref{fig:dme_rot}).  Similarly, in Figure\,\ref{fig:LTE_SYN_DME}, we attempt fits to the data with synthetic spectra of the CH$_3$OCH$_3$ EE lines for different $T_\mathrm{ex}$ values. We vary the column density so as to best-match the observed intensity for the $3_{1,2}-3_{0,3}$ EE transition. For these synthetic spectra fits we assume a fixed line width (0.7\,km/s) and $f = 1$.  {And, while the $4_{1,4}-3_{0,3}$ transition was not detected, we use this as an upper limit anchor. In either method, if one assumes the $5_{1,4}-5_{0,5}$ EE transition is a detection, then a $T_\mathrm{ex}$ of $\sim$30\,K best matches the observations. However, if it is just an upper limit, then the  $T_\mathrm{ex}$ would be lowered ($\sim$10\,K). }

\end{appendix}

\end{document}